\documentclass{jfm}
\usepackage{graphicx}
\usepackage{epstopdf, epsfig}

\usepackage{amsmath, calrsfs, wasysym, verbatim, bbm, color, graphics, physics, bm, bbold, mathtools}
\usepackage{braket}
\usepackage[table]{xcolor}

\newcommand{\VV}{\mathcal{V}}
\newcommand{\one}{\mathbb{1}}

\usepackage{csquotes}

\usepackage{calligra}
\DeclareMathAlphabet{\mathcalligra}{T1}{calligra}{m}{n}
\DeclareFontShape{T1}{calligra}{m}{n}{<->s*[2.2]callig15}{}
\newcommand{\scripty}[1]{\ensuremath{\mathcalligra{#1}}}
\newcommand{\sphr}{\scripty{r}}

\usepackage{booktabs}
\usepackage{array}
\usepackage{makecell}
\usepackage{multirow}
\newcolumntype{L}[1]{>{\raggedright\let\newline\\\arraybackslash\hspace{0pt}}m{#1}}
\newcolumntype{C}[1]{>{\centering\let\newline\\\arraybackslash\hspace{0pt}}m{#1}}
\newcolumntype{R}[1]{>{\raggedleft\let\newline\\\arraybackslash\hspace{0pt}}m{#1}}

\usepackage{hyperref}
\usepackage{bookmark}
\hypersetup{
    colorlinks=true,
    citecolor=blue,      
    linkcolor=blue,
}

\title{Stokes flows in three-dimensional fluids with odd and parity-violating viscosities}

\author{Tali Khain\aff{1,2},
  Colin Scheibner\aff{1,2},
    Michel Fruchart\aff{1,2},
 \and Vincenzo Vitelli\aff{1,2,3}}

\affiliation{
\aff{1}James Franck Institute, The University of Chicago, Chicago, IL 60637, USA
\aff{2}Department of Physics, The University of Chicago, Chicago, IL 60637, USA
\aff{3}Kadanoff Center for Theoretical Physics, The University of Chicago, Chicago, IL 60637, USA
}
\begin{document}

\maketitle

\begin{abstract}
The Stokes equation describes the motion of fluids when inertial forces are negligible compared to viscous forces. 
In this article, we explore the consequence of parity-violating and non-dissipative (i.e. odd) viscosities on Stokes flows in three dimensions. Parity-violating viscosities are coefficients of the viscosity tensor that are not invariant under mirror reflections of space, while odd viscosities are those which do not contribute to dissipation of mechanical energy. 
These viscosities can occur in systems ranging from synthetic and biological active fluids to magnetised and rotating fluids.
We first systematically enumerate all possible parity-violating viscosities compatible with cylindrical symmetry, highlighting their connection to potential microscopic realizations. Then,  
 using a combination of analytical and numerical methods, we analyze the effects of parity-violating viscosities on the Stokeslet solution, on the flow past a sphere or a bubble, and on many-particle sedimentation. 
In all the cases we analyze, parity-violating viscosities give rise to an azimuthal flow even when the driving force is parallel to the axis of cylindrical symmetry. 
For a few sedimenting particles, the azimuthal flow bends the trajectories compared to a traditional Stokes flow. For a cloud of particles, the azimuthal flow impedes the transformation of the spherical cloud into a torus and the subsequent breakup into smaller parts that would otherwise occur.
The presence of azimuthal flows in cylindrically symmetric systems (sphere, bubble, cloud of particles) can serve as a probe for parity-violating viscosities in experimental systems.
\end{abstract}

\let\originalvec\vec
\let\vec\bm

\section{Introduction}

An incompressible fluid is described by the Navier-Stokes equations
\begin{equation}
    \rho D_t \vec{v} = \nabla \cdot \bm{\sigma} + \vec{f}
    \qquad
    \text{and}
    \qquad
    \nabla \cdot  \vec{v} = 0
\end{equation}
in which $\vec{v}$ is the velocity field, $\rho$ the density of the fluid, and $D_t = \partial_t + \vec{v} \cdot \nabla$ is the convective derivative.
Surface forces in the fluid are contained in the stress tensor $\bm{\sigma}$, and body forces such as gravity are contained in $\vec{f}$. 
In a Newtonian fluid, the stress tensor 
\begin{align}
    \sigma_{ij} = \sigma_{ij}^\text{h}+\eta_{ijk\ell} \, \partial_\ell v_k. \label{eq:const}
\end{align}
is composed of a hydrostatic stress $\sigma_{ij}^\text{h}$ present even in the undisturbed fluid (in standard fluids, $\sigma^\text{h}_{ij} = - P \delta_{ij}$ where $P$ is the pressure) and of a viscous stress $\eta_{ijk\ell} \partial_\ell v_k$ that arises in response to velocity gradients.

In Stokes flows, the advection term in the Navier-Stokes equation is small compared to the viscous term (at low Reynolds numbers) and can therefore be neglected~\citep{KimKarrila}. Then, the momentum conservation in the fluid reduces to the (transient/unsteady) Stokes equation
\begin{equation}
    \label{stokes_intro}
    \rho \partial_t \vec{v} = \nabla \cdot \bm{\sigma} + \vec{f}.
\end{equation}
Stokes flows are the setting for phenomena ranging from the locomotion of microscopic organisms~\citep{Purcell1977life,Taylor1951,Lapa2014swimming} to microfluidics~\citep{Stone2004} and sedimentation~\citep{Ramaswamy2001,Guazzelli2009,Goldfriend2017screening, Chajwa2019kepler}. 
In usual fluids such as air and water, the viscosity tensor has only two components, the shear viscosity $\mu$ and the bulk viscosity $\zeta$, the latter of which can be ignored in incompressible flows. Hence, the Stokes equation takes the very simple form
\begin{align}
    \rho \partial_t \vec{v} =& - \nabla P + \mu \Delta \vec{v} + \vec{f} \label{eq:transtokes} 
\end{align}
along with $\bm{\nabla} \cdot {\bm v } = 0$ ($\Delta$ is the Laplacian). 
As the Stokes equation is linear, the flow $\vec{v}$ due to an arbitrary force field $\vec{f}$ can be obtained from the Green function of Eq.~(\ref{eq:transtokes}), called the Oseen tensor, or Stokeslet (see below for precise definitions). This point response can be leveraged to 
describe the flow due to a disturbance in the fluid
or describe the hydrodynamic interactions between colloidal particles.

In this article, we consider a class of fluids called parity-violating fluids. In these fluids, parity (i.e. mirror reflection) is broken at the microscopic level, either by the presence of external fields (e.g. a magnetic field) or by internal activity (e.g. microscopic torques). Parity-violating fluids include fluids under rotation~\citep{Yoshinari1956Kinetic}, magnetized plasma~\citep{Chapman1939}, neutral polyatomic gases under a magnetic field~\citep{Korving1967influence}, but also artificial and biological fluids composed of active elements~\citep{Tsai2005chiral,Condiff1964,soni2019odd,Yamauchi2020} or vortices~\citep{wiegmann2014anomalous} as well as quantum fluids describing the flow of electrons in solids under magnetic field~\citep{Berdyugin2019measuring,Bandurin2016}.
As a consequence of parity violation, the viscous response (summarized by the viscosity tensor) is richer than in usual fluids. 
In three-dimensional polyatomic gases subject to a magnetic field~\citep{Beenakker1970magnetic}, two non-dissipative parity-violating viscosities have been measured~\citep{Korving1967influence,Beenakker1970magnetic} (called $\eta_4$ and $\eta_5$ in these papers). In general, even more parity-violating viscosities can exist. 
In Section~\ref{section_classification}, we classify all possible viscous coefficients of three-dimensional fluids with cylindrical symmetry. Our classification is based on two criteria: whether the viscosities violate parity and whether they contribute to energy dissipation in the fluid. 
We provide a summary of the results that can be used without extensive knowledge of group theory, as well as the underlying group-theoretical analysis. 
In Section~\ref{nonsymmetric_stress_tensor}, we discuss the effects of an antisymmetric hydrodynamic stress. 
In Section~\ref{section_stokeslet}, we analyze in detail how the Stokeslet is affected by the presence of the additional parity-violating viscous coefficients. Qualitatively, the most important change is the presence of an azimuthal velocity in the Stokeslet, which normally vanishes.  
These results allow us to describe the flow past an obstacle in Section \ref{section_flow_past_obstacle}, in which we again find the presence of azimuthal flows, even past a sphere and a spherically symmetric bubble. Finally, in Section \ref{section_sedimentation}, we illustrate the large-scale consequences of parity-violating viscosities in the example of the sedimentation of a cloud of particles under gravity.

\section{The viscosity tensor of a parity-violating fluid}
\label{section_classification}

\subsection{Constraints from spatial symmetries}
\label{spatial_symmetries}

In three dimensions, the rank-four viscosity tensor $\eta_{ijk\ell}$ has 81 possible elements.
However, the form of the viscosity tensor is constrained by the symmetries of the fluid it describes. 
For example, the most general form of the viscosity tensor for an isotropic fluid is given by:
\begin{equation}
    \label{isotropic_viscosity_tensor}
    \eta_{i j k \ell} = \zeta \delta_{ij} \delta_{k \ell}  + \mu \left( \delta_{ik} \delta_{j \ell } + \delta_{i\ell } \delta_{jk} - \frac23 \delta_{ij} \delta_{k \ell }   \right) + \eta_{\text{R}} (\delta_{ik} \delta_{j \ell} - \delta_{i \ell} \delta_{jk})
\end{equation} 
which contains just three independent coefficients: the shear viscosity $\mu$, the bulk viscosity $\zeta$, and the rotational viscosity $\eta_{\text{R}}$ \citep{Groot1962non}. 
These three coefficients are invariant under parity: the exact same coefficients describe the evolution of a fluid and the image of the fluid in a mirror. In an anisotropic fluid, however, this need not be the case.

\begin{figure}
    \centering
    \includegraphics[width=\textwidth]{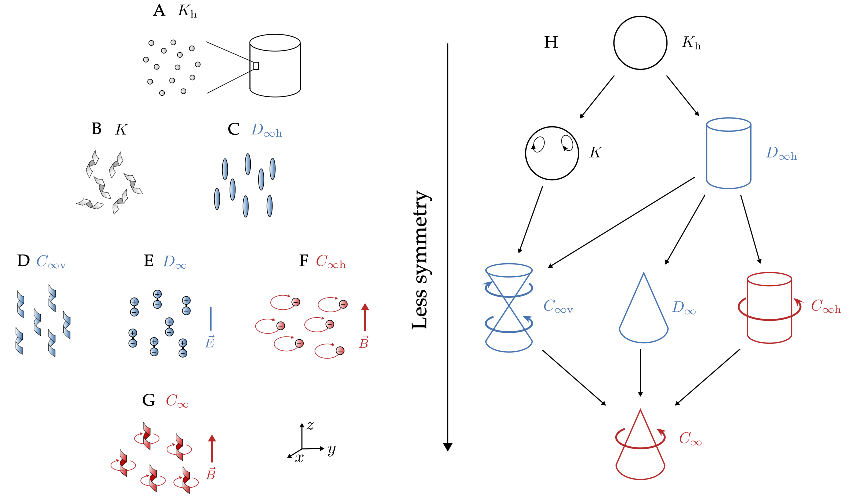}
    \caption{\label{fig:descent_symmetry} 
    Axial symmetry groups, examples of their microscopic realizations, and their constraints on the viscosity tensor. Panels A-G show an example of microscopic system for each axial point group (with cylindrical symmetry about the $\bm{\hat{z}}$ axis) in panel H. Each example is distinguished from the others by the presence of or absence of additional spatial symmetries. 
    (A) A fluid of spherical particles is invariant under all rotations and reflections.
    (B) A fluid of randomly oriented helices (with fixed chirality) is invariant under all rotations, but no reflections. 
    (C) A fluid of elongated (nematic) particles that align with each other is invariant under reflections across all planes parallel and perpendicular to the $\bm{\hat{z}}$ axis. 
    (D) A fluid of chiral particles that align is invariant under $\pi/2$ rotations about any axis perpendicular to the $\bm{\hat{z}}$ axis, but not any reflections.  
    (E) A fluid of electric dipoles under an electric field is invariant under reflections across all planes parallel, but not perpendicular, the $\bm{\hat{z}}$ axis. 
    (F) A fluid of charged particles under a magnetic field (or a fluid of active particle rotating about a fixed axis) is invariant under reflections across all planes perpendicular, but not parallel, to the $\bm{\hat{z}}$ axis.
    (G) A fluid of chiral particles that rotate about a fixed axis has no additional symmetry beyond cylindrical. 
    The group-subgroup relations between axial point groups are shown by arrows in (H). Groups drawn in identical color place identical constraints on the viscosity tensor. 
    The groups $K_h \equiv O(3)$ and $K\equiv SO(3)$ (in black) give rise to the viscosity tensor of an isotropic fluid in Eq.~(\ref{isotropic_viscosity_tensor}). The groups $D_{\infty h}$, $C_{\infty v}$, $D_\infty$ (in blue) allow all the coefficients in black in Eq.~(\ref{eq:planarchiral}) and Table~\ref{viscosity_classification_table}. Some of these coefficients are anisotropic, and all are invariant under reflections parallel and perpendicular to the $\bm{\hat{z}}$ axis (even though the microscopic components are not necessarily invariant under such reflections). The groups $C_{\infty h}$ and $C_\infty$ allow for additional coefficients that change sign under reflection across planes containing the $\bm{\hat{z}}$ axis. These coefficients are shown in red in Eq.~(\ref{eq:planarchiral}) and Table~\ref{viscosity_classification_table}. 
    For more details on the symmetry groups, see~\cite{Shubnikov1988} and \cite{ITA} (in particular Table~\S~10.1.4.2 p.~799 and Fig.~\S~10.1.4.3 p.~803).
    }
\end{figure}

To systematically classify all the viscosity coefficients compatible with a given set of symmetries, we use the language of group theory. A general introduction to group theory in the context of fluid mechanics and applied mathematics is given in \cite{Cantwell2002} and \cite{Hydon2000}. Readers unfamiliar with this formalism can skip directly to Eq.~\ref{eq:planarchiral}, which generalizes the expression in Eq.~\ref{isotropic_viscosity_tensor}.
Figure~\ref{fig:descent_symmetry} and Table~\ref{viscosity_classification_table} provide a visual summary of the possible symmetries of the fluid illustrated by microscopic examples, along with the allowed entries in the viscosity tensor for each symmetry class. 
In general, the less symmetry the fluid has (moving down Fig.~\ref{fig:descent_symmetry}), the larger the number of independent viscosity coefficients. 
Our symmetry analysis can also be read as a guide on how to build parity-violating fluids from microscopic constituents.
The symmetry of the fluid can be designed using the interplay between the symmetries of the microscopic constituents and the way these constituents are collectively arranged in the fluid (for instance, whether they are aligned), see Fig.~\ref{fig:descent_symmetry}A-G and  accompanying caption for concrete examples.

We begin by noting that under a rotation or reflection of space, the viscosity tensor transforms as
\begin{equation}
    \label{viscosity_tensor_transformation}
    \eta_{i j k \ell} = R_{i i'} R_{j j'} R_{k k'} R_{\ell \ell'} \eta_{i' j' k' \ell'}
\end{equation}
where $R$ is an orthogonal matrix that implements the transformation.  
We say a fluid is parity-violating if its properties are not invariant under some improper rotation, i.e., a rotation combined with a reflection. 
In three dimensions, the most general viscosity tensor invariant under all proper rotations [i.e. under the group $SO(3)$, consisting of the transformations $R \in O(3)$ with $\det(R) = 1$] is automatically invariant under all improper rotations as well [i.e. under the whole group $O(3)$]. This happens because any improper rotation can be written as a proper rotation times $-\one = \operatorname{diag}(-1,-1,-1)$: the four copies of $- \one$ always cancel out of Eq.~\ref{viscosity_tensor_transformation}. 

Hence, we have to consider anisotropic fluids in order to see the effects of parity violation.
Here, we focus on systems with cylindrical symmetry (i.e., those invariant under rotation about a fixed axis $\bm{\hat z}$).
The set of all reflections and rotations that leave a fluid globally unchanged forms a group $G$.
It turns out that there are just nine possible symmetry groups that respect cylindrical symmetry~\citep{Shubnikov1988,ITA}. These groups, known as the axial point groups, are shown in 
Figure~\ref{fig:descent_symmetry} and differ from each other by which combinations of horizontal and/or vertical reflections 
are present (see Appendix~\ref{app_symmetries}, in particular Figure~\ref{symmetry_ops}).
Just as invariance under $O(3)$ and $SO(3)$ placed identical constraints on the viscosity tensor, some of the anisotropic symmetry groups in Fig.~\ref{fig:descent_symmetry} place identical constraints on the viscosity tensor. They break into two classes, drawn in blue and in red in Fig.~\ref{fig:descent_symmetry}. 
Fluids with the symmetry groups $D_{\infty\text{h}}$, $C_{\infty\text{v}}$, or $D_{\infty}$ (in blue) have an anisotropic viscosity tensor that 
is invariant under all reflections parallel and perpendicular to the ${\bm \hat z}$ axis. We call these fluids parity-preserving cylindrical, and examples include the aligned nematic particles ($D_{\infty \text{h}}$), aligned helices ($C_{\infty v}$), and dipolar molecules in an electric field ($C_{\infty v}$) shown in Fig.~\ref{fig:descent_symmetry}C-E. 
In contrast, fluids with the symmetry groups $C_{\infty\text{h}}$ or $C_{\infty}$ (in red)  allow additional terms in their viscosity tensor. Examples of such fluids shown in Fig.~\ref{fig:descent_symmetry}F-G include spherical charged particles ($C_{\infty \text{h}}$) and chiral charged particles ($C_{\infty }$) in a magnetic field. 
The additional allowed viscosity coefficients acquire a minus sign when reflected across any plane containing the ${\bm \hat z}$ axis. 
We call these fluids parity-violating cylindrical.

It is useful to organize the components of the viscosity tensor by decomposing the stress $\sigma_{ij}$ and velocity gradient 
$\dot{e}_{k \ell} \equiv \partial_\ell v_k$ 
tensors on a basis of $3 \times 3$ matrices $\tau_{ij}^A$ ($A = 1 \dots 9$) corresponding to a decomposition into irreducible representations of the orthogonal group $O(3)$ (see Appendix~\ref{app_symmetries}). In this notation, the viscosity tensor $\eta_{ijk\ell}$ is expressed as a $9 \times 9$ matrix (see \citet{scheibner2020non, scheibner2020odd}, in which this notation is also used to describe elastic and viscoelastic media).
The basis consists of
\begin{itemize}
    \item a diagonal matrix $\tau_{i j}^1 = C_{ij} = \sqrt{\frac{2}{3}}\delta_{ij}$ corresponding to pressure and dilation,
    \item three anti-symmetric matrices $\tau_{i j}^{A+1} = R_{ij}^A = \epsilon_{Aij}$ corresponding to torques and vorticity,
    \item five traceless symmetric matrices $\tau_{i j}^{A+5} = S^A_{ij}$ corresponding to shear stresses and shear strain rates, whose expressions are
\end{itemize}
\begin{equation*}
\resizebox{.999\hsize}{!}{
    $S^1 =
    \begin{bmatrix}
        1 & 0 & 0 \\
        0 & -1 & 0 \\
        0 & 0 & 0 \\
    \end{bmatrix}
    S^2 =
    \begin{bmatrix}
        0 & 1 & 0 \\
        1 & 0 & 0 \\
        0 & 0 & 0 \\
    \end{bmatrix}
    S^3 =
    \begin{bmatrix}
        \frac{-1}{\sqrt{3}} & 0 & 0 \\
        0 & \frac{-1}{\sqrt{3}} & 0 \\
        0 & 0 & \frac{2}{\sqrt{3}} \\
    \end{bmatrix}
    S^4 =
    \begin{bmatrix}
        0 & 0 & 0 \\
        0 & 0 & 1 \\
        0 & 1 & 0 \\
    \end{bmatrix}
    S^5 =
    \begin{bmatrix}
        0 & 0 & 1 \\
        0 & 0 & 0 \\
        1 & 0 & 0 \\
    \end{bmatrix}
    $}
\end{equation*}
Note that $\tau_{ij}^A \tau_{ij}^B = 2 \delta^{AB}$. 
Defining
\begin{equation}
    \label{irrep_components_from_rank_two_tensors}
    \sigma^A \equiv \sigma_{ij} \; \tau_{ij}^A
    \qquad
    \dot{e}^A \equiv \dot{e}_{i j} \; \tau_{ij}^A
    \qquad
    \eta^{AB} = \frac{1}{2} \; \tau_{ij}^A \, \eta_{ijk\ell} \, \tau_{k\ell}^B
\end{equation}
we may write
\begin{equation}
    \label{viscosity_matrix_def}
    \sigma^A = \eta^{AB} \, \dot{e}^B
\end{equation}
We can transform back to Cartesian tensors via 
\begin{equation}
    \sigma_{ij} =  \frac{1}{2} \; \sigma^A \tau_{ij}^A
    \qquad
    \dot{e}_{ij} =  \frac{1}{2} \; \dot{e}^A \tau_{ij}^A
    \qquad
    \eta_{ijk\ell} = \frac{1}{2} \; \tau_{ij}^A \, \eta^{A B} \, \tau_{k\ell}^B.
\end{equation}
The most general form of $\eta^{A B}$ satisfying cylindrical symmetry about the $\bm{\hat{z}}$ axis is
\begin{equation}
    \includegraphics[width=135mm]{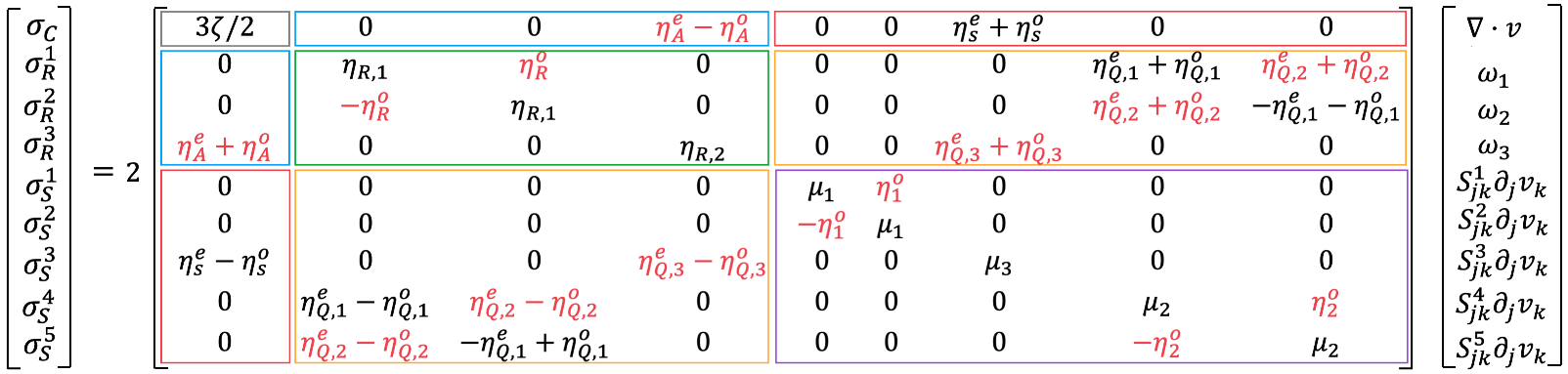} \label{eq:planarchiral}
\end{equation}
in which the parity-violating viscosities are written in red (these are only allowed in the groups drawn in red in Fig.~\ref{fig:descent_symmetry}). An explicit list of parity-violating viscosities is also given in the caption of Table~\ref{viscosity_classification_table}.
Concretely, these entries of the viscosity tensor relate components of the strain rate and stress tensors with different parities under a reflection by a mirror plane containing the $\hat{z}$ axis (see Table \ref{parity_table} for the parities of the basis tensors used in Eq.~\ref{eq:planarchiral} under the reflection $P_y$). Finally, we have restricted our attention to fluids invariant under continuous rotations about the $\bm{\hat{z}}$ axis, because they arise when an originally isotropic fluid is submitted to a single external field.
In general, the fluid can be even less symmetric, for example when the fluid is invariant under a \emph{discrete} point group. This can happen when multiple external fields that are not parallel to each other are applied, or in electron fluids in crystals~\citep{Rao2020,Varnavides2020,Toshio2020,Cook2019}.

\begin{table}
    \centering
    \begin{tabular}{C{0.8cm}C{0.8cm}C{0.8cm}C{0.8cm}C{0.8cm}C{0.8cm}C{0.8cm}C{0.8cm}C{0.8cm}C{0.8cm}}
    \toprule
   & $\sigma_{\text{R}}^1$ & $\sigma_{\text{R}}^2$ & $\sigma_{\text{R}}^3$ & $\sigma_{\text{C}}$ & $\sigma_{\text{S}}^1$ & $\sigma_{\text{S}}^2$ & $\sigma_{\text{S}}^3$ & $\sigma_{\text{S}}^4$ & $\sigma_{\text{S}}^5$ \\
   & $\omega_1$ & $\omega_2$ & $\omega_3$ & $\nabla \cdot v$ & $\dot{e}_{\text{S}}^1$ & $\dot{e}_{\text{S}}^2$ & $\dot{e}_{\text{S}}^3$ & $\dot{e}_{\text{S}}^4$ & $\dot{e}_{\text{S}}^5$ \\
    \midrule
    $P_y$ & $-1$ & $1$ & $-1$ & $1$ & $1$ & $-1$ & $1$ & $-1$ & $1$ \\
    \bottomrule
    \end{tabular}
    \caption{\label{parity_table}
    Effect of the reflection $P_y$ on the components of the stress and strain rate used in Eq.~\ref{eq:planarchiral}.
    The components with a $1$ are invariant under $P_y$, while those with a $-1$ change sign.
    The action of $P_y$ on Cartesian coordinates is $\operatorname{diag}(1,-1,1)$.
    }
\end{table}

\begin{table}
    \centering
    \definecolor{tab_red}{HTML}{D62728}
    \def\pvv#1{\textcolor{tab_red}{#1}}
    \newcolumntype{M}{>{$}c<{$}}
    \begin{tabular}{cccc}
    \toprule
    \makecell{viscosity tensor \\ class} & isotropic & \makecell{parity-preserving \\ cylindrical} & \makecell{ parity-violating \\ cylindrical} \\
    \midrule
    symmetry groups & O(3) \, SO(3) & $D_{\infty h}$ \, $C_{\infty v}$ \, $D_{\infty}$ & $C_{\infty h}$ \, $C_{\infty}$ \\
    \midrule
    \makecell{ dissipative \\ viscosities} & 
    \begin{tabular}{M}
        \eta_{\text{R},1} = \eta_{\text{R},2} \equiv \eta_{\text{R}} \\ 
        \zeta \\
        \mu_1=\mu_2 = \mu_3 \equiv \mu \\
        \\
        \\
    \end{tabular}
    &
    \begin{tabular}{M}
    \eta_{\text{R},1} \; \eta_{\text{R},2} \\
    \zeta  \\ 
    \mu_1 \;\; \mu_2 \;\; \mu_3 \\
    \eta_s^e \\
    \eta_{Q,1}^e
    \end{tabular}
    &
    \begin{tabular}{M}
    \eta_{\text{R},1} \; \eta_{\text{R},2} \\
    \zeta  \\ 
    \mu_1 \;\; \mu_2 \;\; \mu_3 \\
    \eta_s^e \;\; \pvv{\eta_A^{e}}  \\
    \eta_{Q,1}^e \;\; \pvv{\eta_{Q,2}^e} \;\; \pvv{\eta_{Q,3}^e}
    \end{tabular}
    \\
    \midrule
    \makecell{ non-dissipative \\ viscosities} &  
    (none)
    &
    \begin{tabular}{M}
    \\
    \\
    \eta_s^{o} \\
    \eta_{Q,1}^o
    \end{tabular} 
    &
    \begin{tabular}{M}
    \pvv{\eta_\text{R}^o} 
    \\
    \pvv{\eta_1^o} \;\;  \pvv{\eta_2^o}
    \\
    \eta_s^{o} \;\; \pvv{\eta_A^{o}}
    \\ 
    \eta_{Q,1}^o \;\; \pvv{\eta_{Q,2}^o} \;\; \pvv{\eta_{Q,3}^o}
    \\
    \end{tabular} 
    \\
    \bottomrule
    \end{tabular}
    \caption{\label{viscosity_classification_table}
    Classes of viscosity tensors and allowed viscosity coefficients.
    The coefficients refer to Eq.~\ref{eq:planarchiral}.
    Parity-violating viscosities (those only present in the last column) are highlighted in \pvv{red}.
    Explicitly, these are ${\eta_A^{e}}$, ${\eta_{Q,2}^e}$, ${\eta_{Q,3}^e}$, ${\eta_\text{R}^o}$, ${\eta_1^o}$, ${\eta_2^o}$, ${\eta_A^{o}}$,  ${\eta_{Q,2}^o} $,  ${\eta_{Q,3}^o}$.
    See \cite{ITA} for more details on the symmetry groups.
    }
\end{table}

\subsection{Dissipative and non-dissipative viscosities}

In addition to the decomposition based on spatial symmetries discussed in Section~\ref{spatial_symmetries}, the viscosity tensor can be decomposed into symmetric and antisymmetric parts
\begin{align}
    \eta_{ijk\ell} = \eta^{\text{e}}_{ijk\ell} + \eta^{\text{o}}_{ijk\ell}
\end{align}
in which e/o (standing for even/odd) label the symmetric and antisymmetric parts of the tensor, satisfying $\eta^{\text{o}}_{ijk\ell} = - \eta^{\text{o}}_{k\ell ij}$ and $\eta^{\text{e}}_{ijk\ell} = \eta^{\text{e}}_{k\ell ij}$.
The rate of mechanical energy lost by the fluid due to viscous dissipation is (see Appendix~\ref{app:energy}) 
\begin{equation}
    \label{viscous_dissipation_rate}
    \dot{w} = \sigma_{ij} \, \partial_j v_i = \eta_{i j k \ell} \; (\partial_j v_i) \; (\partial_\ell v_k) = \frac{1}{2} \, \eta^{A B} \dot{e}^A \dot{e}^B.
\end{equation}
Hence, the antisymmetric part $\eta^{\text{o}}_{ijk\ell}$ is purely non-dissipative, because $\eta^{\text{o}}_{ijk\ell} (\partial_j v_i) \; (\partial_\ell v_k) = 0$.
In contrast, the symmetric part $\eta^{\text{e}}_{ijk\ell}$ does indeed contribute to viscous dissipation. 
In a standard fluid, the viscous dissipation corresponds to a rate of entropy production $\dot s =  (1/T) \, \sigma_{ij} \partial_j v_i$, where $T$ is temperature. The symmetry of the viscosity tensor has also been related to Onsager reciprocity relations in equilibrium fluids, in which one expects $\eta^{\text{o}}_{ijk\ell} = 0$ when microscopic reversibility is satisfied~\citep{Onsager1931reciprocal,Groot1962non,deGroot1954extension}. 

The dissipative part $\eta^{\text{e}}_{ijk\ell}$ of the viscosity tensor corresponds to the symmetric part of the matrix $\eta^{A B}$ in Eq.~\eqref{viscosity_matrix_def}, while the non-dissipative part $\eta^{\text{o}}_{ijk\ell}$ corresponds to its antisymmetric part. Hence, we have split all off-diagonal terms in Eq.~\eqref{eq:planarchiral} into odd and even parts (except when one of these is already ruled out by spatial symmetry). The non-dissipative viscosities all have a \enquote{o} superscript. In Table~\ref{viscosity_classification_table}, we classify the viscosity coefficients in Eq.~\ref{eq:planarchiral} based on whether they are dissipative or not, and on the symmetry groups in which they can occur.

\section{The stress tensor of a parity-violating fluid}
\label{nonsymmetric_stress_tensor}

In parity-violating fluids, it is possible that the stress tensor is asymmetric. 
An asymmetric stress tensor means that the fluid experiences torques. 
While this is not possible for classical particles interacting through central pairwise interactions, non-central pairwise interactions are sufficient to contribute an antisymmetric part to the stress tensor \citep{Condiff1964}. 
This occurs, for instance, in polyatomic gases since the particles are not spherical~\citep{Condiff1964}. 
In general, anisotropic fluids and fluids with non-symmetric stress require additional hydrodynamic fields, such as the average alignment or angular velocity of the constituents~\citep{Ariman1973,Ramkissoon1976,Hayakawa2000}.  
Here, we assume that all other order parameters relax much faster than the velocity field, so that their dynamics can safely be ignored. 
When the stress tensor is constrained to be symmetric, the viscosity has the additional symmetry $\eta_{ijk\ell} = \eta_{j i k \ell}$.
(Similarly, we have $\eta_{ijk\ell} = \eta_{i j \ell k}$ when vorticity does not affect the viscous response.)

In addition to the viscous stresses discussed in the previous section, the stress tensor also contains a hydrostatic part $\sigma_{ij}^\text{h}$ present even when there is no velocity gradient. Under the assumption of cylindrical symmetry, the hydrostatic stress takes the form
\begin{align}
    \sigma_{ij}^\text{h} = - P \delta_{ij} + \gamma S^3_{ij} - \tau_z R^3_{ij}
\end{align}
in which $P$ is the pressure, $\gamma$ is an hydrostatic shear stress, and $\tau_z$ is a hydrostatic torque. In this paper, we assume that  $\gamma$ and $\tau_z$ are frozen (i.e. they relax to a constant value on very short time scales), like in \cite{banerjee2017odd,markovich2020odd,han2020statistical}. In addition, we assume that $\tau_z$ and $\gamma$ are spatially uniform. In this case, they do not contribute to the term $\partial_j \sigma_{ij}$ in the Stokes Eq.~\eqref{stokes_intro}, and therefore do not affect the form of the Stokeslet, which we discuss in the next section. However, a constant hydrostatic torque $\sigma_{ij}^\text{h} = - \epsilon_{ijk} \tau_k $ can  induce a net torque $T_k$ on an object immersed in the fluid:
\begin{align} 
\label{eq:net_torque}
T_k = \oint_{\partial \VV} \hat n_i \, \sigma_{ji}^h \, \epsilon_{jk\ell} \, x_\ell \, \dd^2 x  = 2 \tau_k \int_\VV \dd^3 x  =2  \tau_k V
\end{align}
where $V$ is the volume of the object $\VV$, in which we have assumed that  $\hat n_i \sigma_{ki} $ is the force on a unit area with normal $\hat n_i$ (this boundary condition might not always hold true, depending on the microscopic interactions and on the definition of the stress).  
The effect of the hydrostatic torque $\tau_z R^3_{ij}$ on a sphere will be further discussed in Sec.~\ref{sec:rotsphere}.
Similarly, the effect of a homogeneous shear stress $\gamma S_{ij}^3$ is to shear a soft deformable body, although it has no effect on rigid bodies.

\section{The Stokeslet of a parity-violating fluid}
\label{section_stokeslet}

\subsection{Oseen tensor and Stokeslet}

The (transient) Stokes equation for an incompressible fluid found in Eq.~\eqref{stokes_intro} can be written as
\begin{equation}
    \label{eq:stokes}
    \rho \partial_t v_i = - \partial_i P + \partial_j [\eta_{i j k \ell} \partial_\ell v_k] + f_i
    \quad
    \text{with}
    \quad
    \partial_i v_i = 0.
\end{equation}
in which we have used the expression \eqref{eq:const} of the viscous stress.
In reciprocal space (see Appendix~\ref{app_coordinate_systems} for Fourier transform conventions),
\def\ii{\text{i}}
\begin{equation}
    \label{stokes_fft}
    -\ii \omega \rho v_i = - \ii q_i P - q_j q_\ell \eta_{i j k \ell} v_k + f_i
    \quad
    \text{with}
    \quad
    \ii q_i v_i = 0.
\end{equation}
These equations can be written as
\begin{equation}
    \label{stokes_M_eq}
    M(\vec{q}, \omega) \vec{v} = - \ii P \vec{q} + \vec{f}
    \quad
    \text{with}
    \quad
    \vec{q} \cdot \vec{v} = 0
\end{equation}
in which we have defined the matrix
\begin{equation}
    \label{M_matrix}
    M_{i k}(\vec{q}, \omega) = q_j q_\ell \eta_{i j k \ell} - \ii \omega \rho \, \delta_{i k}.
\end{equation}
The matrix $M({\bm q}, \omega)$ is always invertible at finite ${\bm q}$ provided that the dissipation rate $\dot{w}$ in Eq.~\ref{viscous_dissipation_rate} is strictly positive (see Appendix~\ref{app:energy}). 
Under this hypothesis, we apply $M^{-1}(\vec{q}, \omega)$ to Eq.~\eqref{stokes_M_eq}. We then take the scalar product with $\vec{q}$ to obtain the pressure $P$, and then replace $P$ with its expression to obtain the velocity, giving
\begin{equation}
    \label{eq:vpfourier}
    \ii P = \frac{\vec{q} \cdot (M^{-1} \vec{f})}{{\vec{q} \cdot (M^{-1} \vec{q})}}
    \quad
    \text{and}
    \quad
    \vec{v} = - \frac{{\vec{q} \cdot (M^{-1} \vec{f})}}{{\vec{q} \cdot (M^{-1} \vec{q})}} \; M^{-1} \vec{q} + M^{-1} \vec{f}.
\end{equation}
The expression of the velocity in terms of the force is then
\begin{equation}
    \vec{v} = G(\vec{q}, \omega) \vec{f}
\end{equation}
in which
\begin{align}
    G_{ij} (\bm{q}, \omega) \equiv &\bigg( [M^{-1}]_{ij}  - \frac{  [M^{-1}]_{im}q_m q_n [M^{-1}]_{nj} }{q_k [M^{-1}]_{k\ell} q_\ell } \bigg ) \label{eq:gen}
\end{align}
is the Green function of the Stokes equation, which is usually called the (reciprocal space) Oseen tensor~\citep{KimKarrila,Kuiken1996}.
Formally, it is defined so that $v_{i} = G_{i j}$ is a solution of Eq.~\eqref{eq:stokes} with $\bm{f} = \delta(\bm{x}) \bm{e}_j$, where $\bm{e}_j$ is the unit vector in direction $j$.
For an isotropic incompressible fluid, we recover the usual (reciprocal space) Oseen tensor 
\begin{equation}
G_{ij}^{\text{iso}}(\vec{q}, \omega=0) = \frac{1}{\mu q^2} \left(\delta_{ij} - \frac{q_i q_j}{q^2} \right).
\end{equation}
When the symmetric part of $\eta_{ijk\ell}$ (corresponding to dissipation) vanishes, the second term of Eq.~\ref{eq:gen} diverges at $\omega = 0$ (but finite $\vec{q}$) because $M^{-1}_{k \ell}$ is strictly anti-symmetric under exchange of $k$ and $\ell$, while the product $q_k q_\ell$ is symmetric (so the denominator $q_k [M^{-1}]_{k\ell} q_\ell$ vanishes). 
This corresponds to a divergence of the characteristic timescale associated with viscous relaxation: in this case, the Stokes approximation is not valid.
In the following, we will assume that viscous relaxation is fast enough, and focus on steady solutions that correspond to the steady Oseen tensor $G(\vec{q}) \equiv G(\vec{q}, \omega = 0)$.

The real space Oseen tensor is then
\begin{equation}
    G_{i j}(\bm{x}) = \frac{1}{(2\pi)^3}\int e^{i \bm q \cdot \bm x} \, G_{i j}(\bm{q}) \, \dd^3 q \label{eq:realspace}
\end{equation}
and the flow generated by a point force $\vec{f}(\vec{x}) =  \vec{f} \delta(\vec{x})$ is the Stokeslet 
\begin{equation}
    \vec{v}(\vec{x}) = G(\bm{x}) \vec{f}
\end{equation}

When the antisymmetric (non-dissipative) part of the viscosity tensor vanishes ($\eta_{ijk\ell}^o = 0$), the matrix $M$ defined by Eq.~\ref{M_matrix} is symmetric. Hence, $M^{-1}$ and the Green function $G(\bm{q})$ are symmetric as well. 
The symmetry of the reciprocal-space Green function $G_{i j}(\bm{q}) = G_{j i}(\bm{q})$ is equivalent to 
 $G_{i j}(\bm{x}) = G_{j i}(\bm{x})$ in physical space. 
This is the expression of Lorentz reciprocity~\citep[\S~4.2, Eq.~4.7]{masoud2019reciprocal}, which can be interpreted as a symmetry in the exchange between the source (producing a force) and the receiver (measuring the velocity field). 
Conversely, Lorentz reciprocity is broken by the presence of non-dissipative (or, equivalently, odd) viscosities.

We can now analyze the effect of parity-violating viscosities on the Stokeslet.
Unlike the situation in a two-dimensional, isotropic incompressible fluid (see Appendix~\ref{app:2D}), in three dimensions the odd and parity-violating viscosities can modify the Stokeslet velocity field.
To see this, we will compute the real-space Oseen tensor or Stokeslet in different cases, using both numerical and analytical methods.
The qualitative changes compared to usual isotropic fluids can be anticipated without any computation from symmetry arguments.
When the driving force is along the axis of azimuthal symmetry, a fluid 
from the classes \enquote{isotropic} and \enquote{parity-preserving cylindrical} in Table~\ref{viscosity_classification_table}
cannot exhibit an azimuthal flow because a reflection symmetry constrains the azimuthal component of the velocity to be opposite to itself -- this is indeed the case for the standard Stokeslet solution~\citep{KimKarrila}.
In contrast, an azimuthal flow is allowed when parity-violating terms are introduced in the viscosity tensor (class \enquote{parity-violating cylindrical} in Table~\ref{viscosity_classification_table}).

\subsection{Stresslet, rotlet, and multipolar responses}
\label{higher_order_responses}

Since the Stokeslet is a response to a point perturbation, multipolar responses can be computed by taking derivatives of the Green function in Eq.~(\ref{eq:realspace}), see~\cite{KimKarrila}.
For example, consider a force dipole defined by a point force $ \vb f$ at $ \frac12 \delta \vb r $ and a point force $-\vb f$ at $-\frac12 \delta \vb r$. The corresponding fluid velocity is given by
\begin{align}
    \label{dipole_G}
    v_i( \vb r) = G_{ik} \qty(\vb r - \frac12 \delta \vb r) f_k - G_{ik} \qty(\vb r + \frac12\delta \vb r) f_k \approx -\partial_j G_{ik} (\vb r) \delta r_j f_k \equiv  H_{ijk} (\vb r) \delta r_j f_k
\end{align}
The tensor $H_{ijk}$ is often decomposed into two contributions: the symmetric part $S_{ijk} = \frac12( H_{ijk} + H_{ikj} )$, which represents the response to point shears (also known as stresslet), and the antisymmetric part $A_{i\ell} = \frac12 \epsilon_{jk \ell} H_{ijk} $ which represents the response to point torques $T_\ell $ (also known as rotlet). As discussed in Sec.~\ref{nonsymmetric_stress_tensor}, such point torques can arise from an hydrostatic torque in the fluid. 
An explicit expression of the Oseen tensor $G_{i k}$ is given by \eqref{perturbative_oseen_tensor} of Appendix~\ref{app:stokeslet} in a perturbative case, from which the stresslet and rotlet can be deduced using Eq.~\eqref{dipole_G}.

\subsection{General numerical solution}
\label{sec:numerics}

To determine the physical-space Stokeslet or Oseen tensor $G(\vec{x})$, one must compute the inverse Fourier transform \eqref{eq:realspace}.
This can be done numerically in the general case, in which analytical solutions are not easily accessible. To do so, we evaluate Eq.~\eqref{eq:gen} on a discrete grid in reciprocal space (each component of ${\bm q}$ ranges from $-Q$ to $+Q$ with increments $\delta q$). This allows us to resolve length scales larger than a few $\pi/Q$ but smaller than $\pi/\delta q$. We then use the fast Fourier transform (FFT) algorithm to compute the real-space Oseen tensor (or the real-space Stokeslet). To avoid numerical instabilities (Gibbs oscillations) due to the sharp cutoff in reciprocal space, we regularize the integrand in Eq.~\eqref{eq:realspace} with a Gaussian kernel $e^{- \pi q^2/ 4 Q^2}$~\citep{Gomez2013,Cortez2001}.
This procedure allows us to compute the Stokeslet for an arbitrary viscosity tensor. Our code for this computation is available at \url{https://github.com/talikhain/StokesletFFT}.

We consider an external force parallel to the ${\bm \hat{z}}$ axis and examine the perturbative effect of each coefficient separately. 
We set the normal shear viscosity to $\mu_1= \mu_2 = \mu_3 = 1$ and vary each of the other viscosity coefficients one by one, setting them to be $\eta_i = 0.01 \mu$. 
This flow is visualized for each viscosity in Fig.~\ref{fig:numerics} of Appendix~\ref{app:numerics}, in which we also validate the numerical method using the exact solution discussed in the next section, see Fig.~\ref{fig:numvalidodd}.
We find that the viscosity coefficients that give rise to an azimuthal flow are
\begin{equation}
\eta_R^o 
\quad
\eta^e_{Q,2}
\quad
\eta^o_{Q,2}
\quad
\eta^e_{Q,3}
\quad
\eta^o_{Q,3}
\quad
\eta_1^o
\quad
\eta_2^o
\end{equation}
The list of viscosity coefficients that we found to generate $v_{\phi}$ are a subset of the parity-violating viscosities (printed in red in Eq.~\eqref{eq:planarchiral} and listed in the caption of Table~\ref{viscosity_classification_table}), as expected. In fact, the only parity-violating viscosities that do not give rise to azimuthal flow are $\eta^e_A$ and $\eta^o_A$. 
This is because we have assumed that the flow is incompressible.
First, the term $(\eta_A^e - \eta_A^o) \nabla \cdot \bm{v}$ vanishes because $\nabla \cdot \bm{v} = 0$.
Second, the term $(\eta_A^e + \eta_A^o) \omega_3$ contributes to the component $\sigma_C$ of the stress, and can therefore be absorbed in the pressure.

\subsection{The Stokeslet of an odd viscous fluid: exact solution}
\label{sec:exact}

\begin{figure}
 \begin{center}
\leavevmode
\includegraphics[width=135mm]{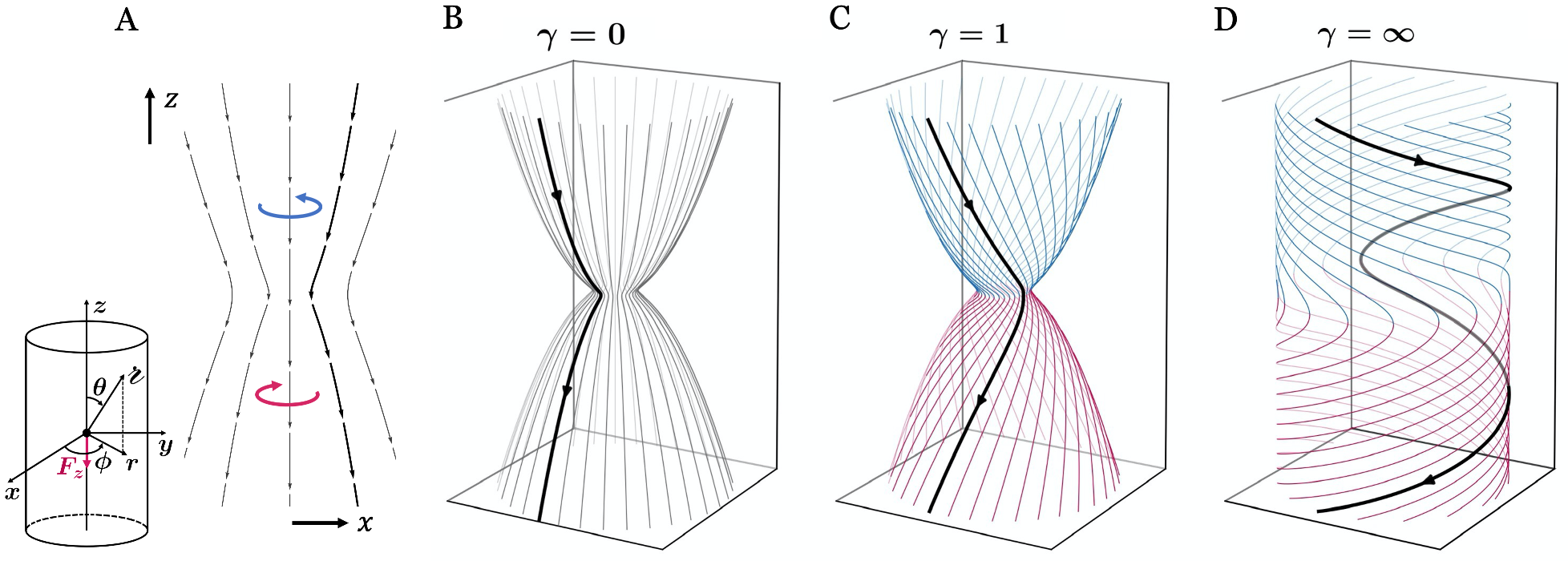} 
\caption{A Stokeslet in an odd viscous fluid. A. The streamlines of a standard Stokeslet flow are shown in black. The blue and red arrows indicate the appearance of an azimuthal flow once odd viscosity is introduced. A schematic of the system and the coordinate convention is found in the inset. An external force, $\bm{F}$, is applied at the origin in the $-\hat{\bm{z}}$ direction. B-D. A three-dimensional rendition of the Stokeslet streamlines initialized around a circle (i.e. many copies of the bolded streamline in panel B), for a range of viscosity ratios, $\gamma = \eta^o/\mu$. As the odd viscosity increases, the velocity field develops an azimuthal component that changes sign across the $z = 0$ plane, where the source is located. In the limit of only odd viscosity (panel D), the familiar radial component of the flow vanishes.}
\label{fig:stokesletexact}
\end{center}
\end{figure}

We now consider a particular case in which the real-space Stokeslet can be computed analytically. 
First, we set $\mu\equiv\mu_1=\mu_2=\mu_3$ and consider only the odd shear viscosities $\eta_1^o$ and $\eta_2^o$ (all the other viscosities are assumed to vanish, except perhaps the bulk $\xi$ viscosity which drops out of the Stokes equation). 
In this case, the matrix $M(\omega = 0)$ defined by Eq.~\ref{M_matrix} takes the form
\begin{align}
    M(\omega = 0) = 
    \begin{bmatrix}
    \mu q^2 & \eta_1^o(q_x^2 + q_y^2) - \eta_2^o q_z^2 & -\eta_2^o q_y q_z  \\
    - \eta_1^o(q_x^2 + q_y^2) + \eta_2^o q_z^2 & \mu q^2 & \eta_2^o q_x q_z  \\
    \eta_2^o q_y q_z & -\eta_2^o q_x q_z  & \mu q^2  \\
\end{bmatrix}
\label{eq:M}
\end{align}

Taking ${\bm f} = -\bm{\hat z} F_z \delta^3({\bm x})$, and defining $q_{\perp}^2 \equiv q_x^2 + q_y^2$, we find the full expressions for the velocity and pressure in Fourier space by using Eq.~\ref{eq:vpfourier},
\begin{align}
\bm{\hat{v}(\bm{q})}
&=
\frac{F_z}{N(\vec{q})}
\begin{bmatrix}
    q_z(q_y(-(\eta_1^o + \eta_2^o)q_{\perp}^2 +\eta_2^o q_z^2) + \mu q_x (q_{\perp}^2 + q_z^2))\\
    \\
    q_z(q_x((\eta_1^o+\eta_2^o)q_{\perp}^2 -\eta_2^o q_z^2) + \mu q_y (q_{\perp}^2 + q_z^2))\\
    \\
     -\mu q_{\perp}^2(q_{\perp}^2 + q_z^2)\\
\end{bmatrix} \label{eq:vfourier_etas}\\
\hat{p}(\bm{q})
&=
i \frac{F_z}{N(\vec{q})} \; q_z[(\eta_1^o q_{\perp}^2 - \eta_2^o q_z^2)((\eta_1^o + \eta_2^o)q_{\perp}^2 - \eta_2^oq_z^2) + \mu^2(q_{\perp}^2 + q_z^2)^2] \label{eq:pfourier_etas}
\end{align} 
in which $N(\vec{q}) = \mu^2(q_{\perp}^2 + q_z^2)^3+q_z^2((\eta_1^o + \eta_2^o)q_{\perp}^2 - \eta_2^o q_z^2)^2$.
Second, we assume that $\eta_1^o = -2 \eta_2^o$, for which simplifications occur in  Eqs.~\ref{eq:vfourier_etas}-\ref{eq:pfourier_etas}.
This particular case occurs in the limit of low magnetic field regime in experiments on polyatomic gases (see Eq.~(13) in \cite{Hulsman1970transverse}), and was also obtained in a theoretical Hamiltonian description of fluids of spinning molecules~\citep{markovich2020odd}.
In this limit, the viscosity matrix can be seen as a simple combination of isotropic contractions and rotations about the $\bm{\hat{z}}$ axis in the space of shears (see Appendix~\ref{app:exact}).

In order to find the real space solution, we compute the inverse Fourier transform in Eq.~\ref{eq:realspace} (see Appendix~\ref{app:exact} for the detailed calculation).
Parameterizing the final flow field by $\gamma = \eta_2^o/ \mu$, we obtain the velocity field
\begin{align}
    v_r(
    \sphr, \theta) &=-\frac{F_z}{4\pi \eta_2^o}\frac{\cot{\theta}}{\gamma\sphr} \left(1 - \frac{1}{ \sqrt{1 + \gamma^2\sin^2\theta}}\right) \label{eq:str}
\end{align}
\begin{align}
    v_{\phi}(\sphr, \theta) &=\frac{F_z}{4\pi \eta_2^o}\frac{\cot{\theta}}{\sphr} \left(1 -\frac{1}{ \sqrt{1+\gamma^2 \sin^2{\theta}}}\right) \label{eq:stphi} \\
    v_z(\sphr, \theta) &=\frac{F_z }{4\pi \eta_2^o  } \frac{1}{ \gamma \sphr}\left(1 -\frac{\gamma^2 + 1}{ \sqrt{1+\gamma^2\sin^2 \theta}}\right)  \label{eq:stz}
\end{align}
as well as the pressure field
\begin{align}    
    p(\sphr,\theta) &=\frac{F_z}{4\pi}\frac{\cos{\theta}}{\sphr^2}\left(1 - \frac{2(\gamma^2 + 1)}{(1 + \gamma^2 \sin^2{\theta})^{3/2}}\right) \label{eq:stp}
\end{align}
Here, $\sphr$ is the radius in spherical coordinates (see schematic in Figure~\ref{fig:stokesletexact}A and Appendix~\ref{app_coordinate_systems}).
Streamlines of the velocity field are visualized for a range of $\gamma$ in Fig.~\ref{fig:stokesletexact} and Supplementary Movie 1. In the absence of odd viscosity, the Stokeslet flow only has two components, $v_r$ and $v_z$ (Appendix~\ref{app:exact}), visualized in the vertical $x$-$z$ plane in Fig.~\ref{fig:stokesletexact}A and in three dimensions in Fig.~\ref{fig:stokesletexact}B. Notably, as the blue and red arrows in Fig.~\ref{fig:stokesletexact}A indicate, the flow develops an azimuthal component for $\gamma \neq 0$ (Fig.~\ref{fig:stokesletexact}C-D), consistent with the fact that $\eta_1^o$ and $\eta_2^o$ are parity-violating (see Table~\ref{viscosity_classification_table} and Eq.~\ref{eq:planarchiral}).  

As $\gamma$ is increased, the magnitude of the azimuthal component grows, while the radial component diminishes. When $\gamma \gg 1$, the $\bm{\hat{r}}$-component of the velocity field goes to zero, while $v_{\phi}$ and $v_z$ approach $(\sphr\sin{\theta})^{-1}$. 
(For the approximation of a steady Stokes flow to remain valid, the dissipative shear viscosity $\mu$ must remain finite in order to ensure that the relaxation time of the fluid is also finite, so the limit $\gamma = \infty$ is never actually reached.) 
At smaller $\gamma$, the central line splits into lobes of high azimuthal velocity that migrate away from the vertical, as illustrated in Appendix~\ref{app:exact}. 

\subsection{Stokeslet: perturbative solution}
\label{sec:stokeslet}

In this section, we consider more generally the effect of the odd shear viscosities and of the rotational viscosities on the Stokeslet by treating the problem perturbatively (with respect to the small parameters characterizing the magnitude of these viscosities). We find that the first order correction $\bm{v}_{\text{Stokes},1}$ to the standard Stokeslet (given in Eq.~\ref{eq:appstokesnorm} of Appendix~\ref{app:exact}) due to the parity-violating coefficients $\eta_1^o, \eta_2^o$ and $\eta_R^o$ is of the form
\begin{align}
    \bm{v}_{\text{Stokes},1} = v_{\phi,1}\bm{\hat{\phi}}
    = \left[ 
    v_{\phi, 1}^{(\eta_1^\text{o})}
    + 
    v_{\phi, 1}^{(\eta_2^\text{o})}
    +
    v_{\phi, 1}^{(\eta_R^o)}
    \right]
    \bm{\hat{\phi}}
\end{align}
Let us now discuss the explicit form of each of these terms, starting with the odd shear viscosities.

\begin{figure}
 \begin{center}
 \leavevmode
\includegraphics[height=50mm]{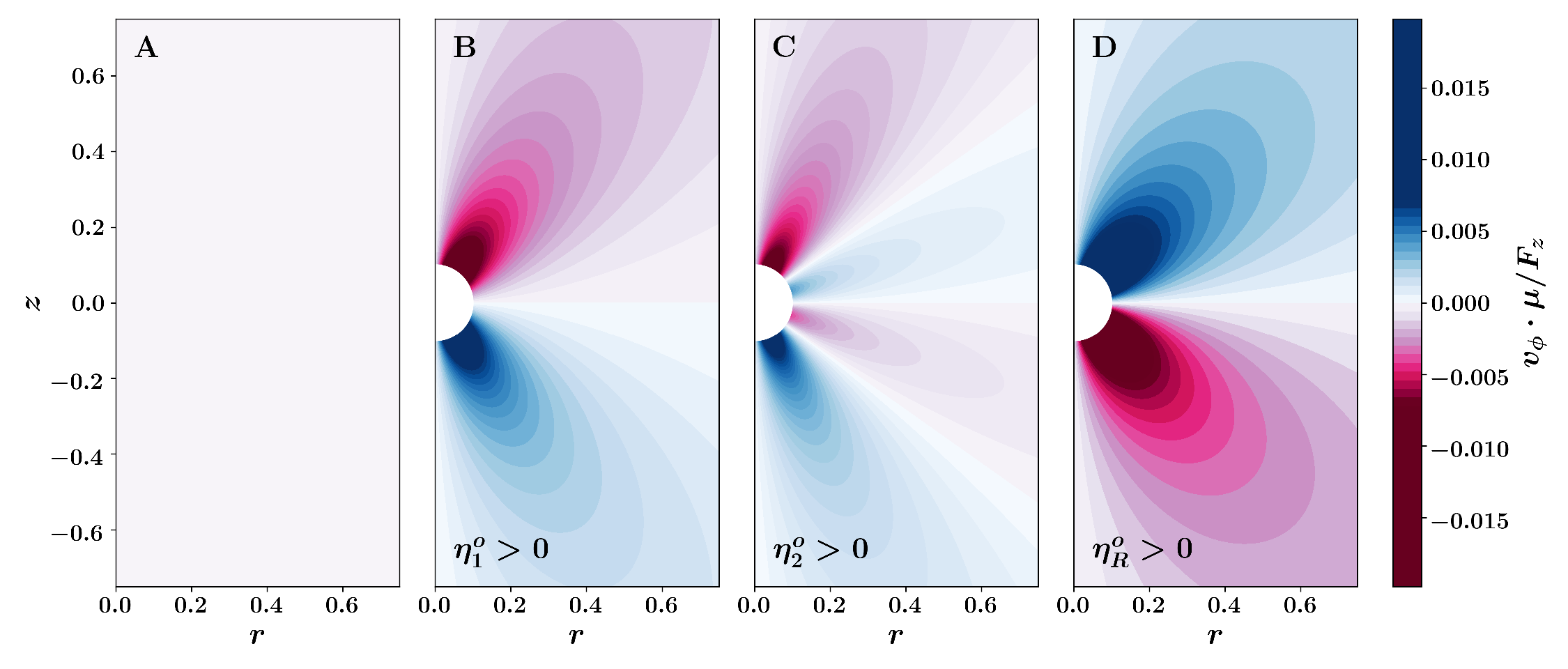} 
\caption{The non-dimensionalized azimuthal component of the Stokeslet flow for small shear and rotational odd viscosity coefficients. A. In the absence of odd viscosity, the azimuthal component of the velocity field is zero. B-D. The first order correction of the Stokeslet due to $\eta_1^o, \eta_2^o$, and $\eta_R^o$, respectively, taking $\eta^o/\mu = 0.1$. The origin is removed due to the singularity of the flow at $r = z = 0$. The azimuthal flow is odd with respect to $z$, and forms lobe-like regions of concentrated rotation. Blue indicates flow into the page, red corresponds to flow out of the page.
Overall, the fluid flows out of the page in the upper lobe (in red) and into the page in the lower lobe (in blue). In C, two small additional lobes have opposite velocities compared to the bigger ones.
}
\label{fig:stokesletpert}
\end{center}
\end{figure}

Starting back from Eqs.~\ref{eq:vfourier_etas}-\ref{eq:pfourier_etas} (in which $\eta_1^o$ and $\eta_2^o$ are independent), we perform a perturbative expansion in the quantities $\epsilon_{1(2)} \equiv \eta_{1(2)}^o/\mu  \ll 1$. Computing the inverse Fourier transform to obtain the flow fields in real space as in Section \ref{sec:exact} (see Appendix~\ref{app:stokeslet} for the detailed calculation), we find that both $\eta_1^o$ and $\eta_2^o$ contribute to leading order by introducing terms contained entirely in the $\bm{\hat{\phi}}$-component of the velocity field.
The contributions of the two viscosities are
\begin{align}
    v_{\phi, 1}^{(\eta_1^\text{o})}(\sphr,\theta)&=-\epsilon_1 \frac{F_z} {128 \pi \mu}\frac{(5 + 3\cos{2\theta})\sin{2\theta}}{\sphr}+\order{\epsilon_1^2} \label{eq:eta1}\\
    v_{\phi, 1}^{(\eta_2^\text{o})}(\sphr, \theta)&=-\epsilon_2 \frac{F_z} {64 \pi \mu } \frac{(1 + 3\cos{2\theta})\sin{2\theta}}{\sphr}+\order{\epsilon_2^2} \label{eq:eta2}
\end{align}
while the presssure is not modified at first order. 
The azimuthal component is visualized in the vertical $r$-$z$ plane in Fig.~\ref{fig:stokesletpert}. In the absence of odd viscosity (Fig.~\ref{fig:stokesletpert}A), $v_{\phi}=0$. The non-dimensionalized $v_{\phi}$ profiles for $\eta_1^o$ and $\eta_2^o$ given by Eq.~\ref{eq:eta1}-\ref{eq:eta2} are shown in Fig.~\ref{fig:stokesletpert}B-C. While both velocity fields decay as $1/\sphr$, they differ appreciably in their angular dependence: $\eta^o_2$ includes an additional sign change. 

We now consider rotational viscosities, which couple vorticity and torques.
These viscosities break both minor symmetries of the viscosity tensor ($\eta_{ijk\ell} \neq \eta_{jik\ell} \neq \eta_{ji\ell k}$), because the vorticity and torques are the antisymmetric parts of the strain rate and stress tensors, and are shown in the block outlined in green in Eq.~\ref{eq:planarchiral}, reproduced below:
\begin{align}
\begin{bmatrix}
    \sigma^1_R\\
    \sigma^2_R\\
    \sigma^3_R\\
\end{bmatrix}
=
\begin{bmatrix}
    \eta_{R,1} && \eta_R^o && 0 \\
    -\eta_R^o && \eta_{R,1}  && 0 \\
    0 && 0 && \eta_{R,2}\\
\end{bmatrix}
\begin{bmatrix}
    \omega_1\\
    \omega_2\\
    \omega_3\\
\end{bmatrix}
\label{eq:block}
\end{align}
These rotational viscosities are often ignored in standard fluids because their contribution to the stress relaxes to zero over short times~\citep{Groot1962non}, but occur in the hydrodynamics of liquid crystals \citep{Miesowicz1946,Leslie1968,Ericksen1961,Parodi1970} as well as in the hydrodynamics of electrons in materials with anisotropic Fermi surfaces~\citep{Cook2019}.

We consider perturbations in the quantities
$\epsilon_{R,1} = \eta_{R,1}/\mu, \epsilon_{R,2} = \eta_{R,2}/\mu$ and $\epsilon_R^o = \eta_R^o/\mu$. 
The matrix ${M}$ is given by
\begin{align}
    {M} = \mu
    \begin{bmatrix}
    q^2 - \epsilon_{R,1} q_z^2 - \epsilon_{R,2}q_y^2 &  \epsilon_{R,2} q_x q_y -\epsilon_R^o q_z^2 & \epsilon_{R,1} q_x q_z + \epsilon_R^o q_y q_z \\
    \epsilon_{R,2} q_x q_y + \epsilon_R^o q_z^2 & q^2 -\epsilon_{R,1} q_z^2 - \epsilon_{R,2}q_x^2 & -\epsilon_R^o q_x q_z + \epsilon_{R,1} q_y q_z \\
    \epsilon_{R,1} q_x q_z - \epsilon_R^o q_y q_z &  \epsilon_R^o q_x q_z + \epsilon_{R,1} q_y q_z&  q^2 - \epsilon_{R,1} q^2  \\
\end{bmatrix}
\end{align}

Applying Eq.~\ref{eq:vpfourier}, we calculate the velocity and pressure in Fourier space,
\begin{align}
\bm{v}(\bm{q})
&=
\frac{F_z}{\mu \; N_2(\vec{q})}
\begin{bmatrix}
     - q_x q_z (q_{\perp}^2 + q_z^2)  -\epsilon_R^o q_y q_z (q_{\perp}^2 + q_z^2) + \epsilon_R q_x q_z^3\\
    \\
    - q_y q_z (q_{\perp}^2 + q_z^2) + \epsilon_R^o q_x q_z (q_{\perp}^2 + q_z^2) + \epsilon_R q_y q_z^3\\
    \\
     q_{\perp}^2(q_{\perp}^2 + q_z^2) - \epsilon_R q_{\perp}^2 q_z^2\\
\end{bmatrix} \label{eq:vfourier_rot}\\
p(\bm{q}) &= \frac{-\ii Fz}{N_2(\vec{q})} \; [q_z (q_{\perp}^2 + q_z^2)^2 -\epsilon_R q_z (q_{\perp}^2 + q_z^2)(q_{\perp}^2 + 2q_z^2)+ (\epsilon_R^2 + (\epsilon_R^o)^2) q_z^3(q_{\perp}^2 + q_z^2)]
\label{eq:pfourier_rot}
\end{align}
in which
\begin{align}
    N_2(\vec{q}) = {(q_{\perp}^2 + q_z^2)^3+ \epsilon_R(q_{\perp}^2 + q_z^2)^2(-q_{\perp}^2 - 2q_z^2) + (\epsilon_R^2 + (\epsilon_R^o)^2)q_z^2 (q_{\perp}^2 + q_z^2)^2}.
    \label{eq:fourier_rot_denom}
\end{align}
Note that the coefficient $\epsilon_{R,2}$ does not affect the flow (see Appendix \ref{app:stokeslet} for further details). For the remaining coefficients, we expand the above expressions up to first order in $\epsilon_{R,1}$ and $\epsilon_R^o$, and compute their inverse Fourier transform to find the real space fields.

Of the three rotational viscosities, only $\eta_R^o$ violates parity (see Table~\ref{viscosity_classification_table} and Eq.~\ref{eq:planarchiral}) and as a consequence, gives rise to an azimuthal flow,
\begin{align}
    v_{\phi, 1}^{(\eta_R^o)}(\sphr,\theta)&=\epsilon_R^o\frac{F_z}{16 \pi \mu}\frac{\sin(2 \theta )}{ \sphr}+\order{(\epsilon_R^o)^2} \label{eq:etaR}
\end{align}
The $v_\phi$ profile due to $\eta_R^o$ is shown in Fig.~\ref{fig:stokesletpert}D.
While the parity-violating shear and rotational viscosities generate quantitatively different azimuthal flows, their qualitative effect is the same. 
The pressure is again not modified to first order.

\section{Odd viscous flow past an obstacle}
\label{section_flow_past_obstacle}

\subsection{Odd viscous flow past a sphere}
\label{sec:sphere}

Two-dimensional flows past obstacles in the presence of a non-dissipative (odd) viscosity have previously been studied experimentally in \citet{soni2019odd} and theoretically in \citet{Kogan2016Lift}.
\citet{Lapa2014swimming} also analyzed the consequences on swimmers at low Reynolds numbers.
In these two-dimensional cases, only the pressure field is modified by the additional viscous terms, while the velocity field remains unchanged. Nevertheless, \citet{Kogan2016Lift} reported that a lift force appears in the Oseen approximation (including inertia) of the flow past an infinite cylinder due to the non-dissipative viscosity. In this section, we consider three-dimensional flows. Even in the Stokes limit (without inertia), we find that parity-violating viscosities have a qualitative effect on the flow past a sphere: the Stokes drag is not modified at this order, but an azimuthal velocity develops despite the symmetry of the obstacle.

Let us begin by considering the viscous flow past a {\it finite} radius sphere~\citep{KimKarrila}. We assume a uniform velocity field $\bm{v} =U \hat {\bm z}$ at $\sphr \to \infty$ and a no-slip boundary condition with $\bm{v}=0$ on the surface of the sphere $\sphr=a$. 
The streamlines of this flow in a standard fluid are shown in black in Fig.~\ref{fig:spherebubble}A on the $r$-$z$ plane.
Here, we assume that the sphere cannot (or does not) rotate. 
In Sec.~\ref{sec:rotsphere}, we will discuss the case in which the sphere is allowed to rotate.

We once again look for a perturbative solution to Eq.~\ref{eq:stokes} with $f_i = 0$ in the small parameters $\epsilon_1$, $\epsilon_2$, and $\epsilon_R^o$. To leading order in the parity-violating viscosities, the pressure field about the sphere is given simply by the pressure term due to the Stokeslet, as in a standard isotropic fluid. Since the Stokeslet pressure does not have a first order correction (Section \ref{sec:stokeslet}), Eq.~\ref{eq:stokes} reduces to the vector Poisson equation for the first order velocity field,
\begin{align}
    \Delta \bm{v_1} = - \Delta_{\alpha} \bm{v_0} \label{eq:vecpoisson}
\end{align}
in which $\Delta_{\alpha}$ is the second-order differential operator associated to the viscosity $\alpha$ (here, $\alpha = \eta_1^\text{o}, \eta_2^\text{o}, \eta_\text{R}^\text{o}$, see Appendix \ref{app:oddop} for explicit form),
and $\bm{v_0}$ is the flow past a sphere in a standard fluid (given by Eq.~\ref{eq:appnormsphere}). The resulting vector Poisson equation for the perturbed flow is formally equivalent to the electrostatics problem of finding the electric potential due to a conducting spherical cavity enclosing a point charge. We use the corresponding Dirichlet Green function by expanding the solution in spherical harmonics \citep{jackson1999classical}. The details of this calculation are provided in Appendix~\ref{app:sphere}.
Solving for the flow $\bm{v_1}$ to leading order in $\epsilon_1$, $\epsilon_2$, and $\epsilon_R^o$, we can express the resulting velocity field in terms of the Stokeslet solution, $\bm{v_{\text{Stokes}, 1}}$ in  Eqs.~\ref{eq:eta1}, \ref{eq:eta2}, and \ref{eq:etaR}, from Section \ref{sec:stokeslet}, as
\begin{align}
    v_{\phi, 1} (\sphr, \theta) &=\left(\frac{6\pi a U \mu}{F_z} \bm{v_{\text{Stokes}, 1}} + \frac{\pi a^3 U \mu}{F_z}\Delta \bm{v_{\text{Stokes}, 1}}  + \frac{\pi a^5 U \mu}{20 F_z}\Delta^2 \bm{v_{\text{Stokes}, 1}} \right)\cdot \bm{\hat{\phi}}
    \label{eq:spherestokeslet}
\end{align}
with no modifications to $v_r$ and $v_z$ at leading order.

In standard isotropic fluids, a superposition of the Stokeslet ($v \propto 1/\sphr$) and its second derivative  (a source dipole $v \propto 1/\sphr^3$) is sufficient to satisfy the boundary conditions. In the presence of odd viscosity, we find that higher order in gradients are necessary, as can be seen from Eq.~\ref{eq:spherestokeslet}. Even so, by equating the far field of the flow and the Stokeslet solution, we find $F_z = 6\pi a U\mu$. Hence, the Stokes drag experienced by the sphere remains unchanged to first order in $\eta_R^o$, $\eta_1^o$, and $\eta_2^o$ compared to a standard fluid.

Rewriting Eq.~\ref{eq:spherestokeslet} more explicitly, we have
\begin{align}
    v_{\phi, 1} (\sphr, \theta) &=\frac{3U}{64}\left[  g^1(\theta) \frac{a}{\sphr} + g^3(\theta) \left(\frac{a}{\sphr} \right)^3 + g^5(\theta) \left(\frac{a}{\sphr} \right)^5 \right] \sin{2\theta}
    \label{eq:sphere}
\end{align}
where
\begin{align}
    g^1(\theta)=&\ 8\epsilon_R^o -  ( 5 + 3\cos{2\theta}  ) \epsilon_1  -  (2 + 6\cos{2\theta}) \epsilon_2 +\order{\epsilon^2} \\
    g^3(\theta)=&-8\epsilon_R^o +  (6 +10\cos{2\theta}) \epsilon_1 + (4 +20\cos{2\theta}) \epsilon_2+\order{\epsilon^2} \\
    g^5(\theta)=&   -( 1 + 7\cos{2\theta}  ) \epsilon_1 - (2 + 14\cos{2\theta}) \epsilon_2 +\order{\epsilon^2}
\end{align}
These velocity fields are shown in Fig.~\ref{fig:spherebubble}C-E on the $r$-$z$ plane. As in the Stokeslet case, each odd viscosity coefficient results in an azimuthal flow, but the quantitative features of the velocity field vary depending on the exact viscosity chosen.

\subsection{Odd viscous flow past a bubble}
\label{sec:bubble}

\begin{figure}
 \begin{center}
 \leavevmode
\includegraphics[width=135mm]{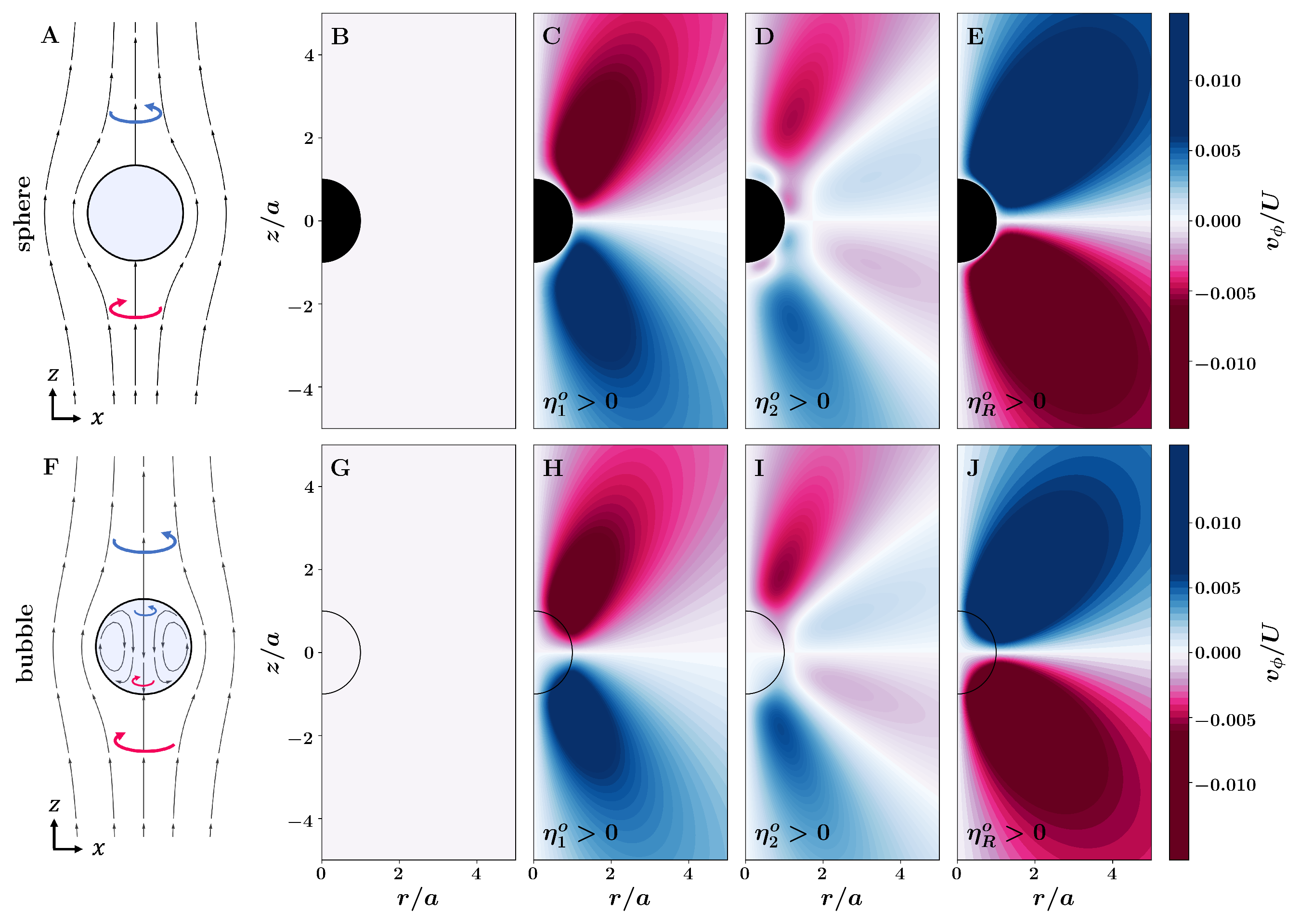} 
\caption{Odd viscous flow past a sphere (panels A-E) and a bubble (panels F-J). A. The streamlines of a standard flow past a sphere are shown in black. The blue and red arrows indicate the appearance of an azimuthal flow once odd viscosity is introduced. B-E. The non-dimensionalized azimuthal velocity component of the flow visualized on the $r$-$z$ plane. If the odd viscosity is absent (panel B), the azimuthal component is zero. Perturbative additions of $\eta_1^o, \eta_2^o$, and $\eta_R^o$ (taking $\eta^o/\mu = 0.1$) significantly affect the flow past a sphere by introducing a nonzero $v_{\phi}$ that is odd in $z$ (panels C-E). F. The streamlines of a standard flow outside and inside a spherical bubble are shown in black. The blue and red arrows indicate the appearance of an azimuthal flow once odd viscosity is introduced. G-J. The non-dimensionalized azimuthal velocity component of the flow visualized on the $r$-$z$ plane. Unlike the case of the sphere, the velocity field extends into the bubble, with a continuous velocity across the bubble surface. If the odd viscosity is absent (panel G), the azimuthal component is zero. Perturbative additions of $\eta_1^o, \eta_2^o$, and $\eta_R^o$ (taking $\eta^o/\mu = 0.1$) significantly affect the flow by introducing a nonzero $v_{\phi}$ both inside and outside the bubble (panels C-E).}
\label{fig:spherebubble}
\end{center}
\end{figure}

Closely related to the flow past a solid sphere is the flow past a spherical bubble without surface tension, in which the bubble itself is filled with a fluid \citep{hadamard1911slow,rybczynski1911uber, lamb1924hydrodynamics, batchelor2000introduction}.
Here we assume that the inner and outer fluid have the same viscosities.
As with the sphere, we solve for a steady velocity field configuration satisfying $\bm v = U \bm{\hat z} $ as $\sphr \to \infty$, but now we require that the velocity field be continuous throughout all space (even across the nominal boundary of the bubble). 
In a standard fluid, the flow outside the bubble resembles that of the flow past a sphere, while the flow inside is described by Hill's spherical vortex \citep{hill1894vi}, with the boundary condition imposing continuous velocity at the surface. The streamlines of this velocity field are visualized in black in Fig.~\ref{fig:spherebubble}F. Following the setup above, let us consider the effect of the odd viscosities $\eta_1^o, \eta_2^o, \eta_R^o$ in the perturbative limit.

Like in the case of the sphere, the first order correction to the pressure vanishes, and the flow outside the bubble reduces to Eq.~\ref{eq:vecpoisson}.
To solve this equation, we again employ Green function methods. Unlike the sphere problem, however, the boundary condition no longer requires no-slip velocity on the bubble surface, so we do not need to use the Dirichlet Green function. The details of this calculation are provided in Appendix~\ref{app:bubble}. Solving for the flow to leading order in $\epsilon_1, \epsilon_2,$ and $\epsilon_R^o$, we can write it in terms of the Stokeslet solution, $\bm{v_{\text{Stokes}, 1}}$,
\begin{align}
    v_{\phi, 1}^{\text{out}} (\sphr, \theta) &=\left(\frac{5\pi a U \mu}{F_z} \bm{v_{\text{Stokes}, 1}} + \frac{\pi a^3 U \mu}{2 F_z}\Delta \bm{v_{\text{Stokes}, 1}}  + \frac{\pi a^5 U \mu}{56 F_z}\Delta^2 \bm{v_{\text{Stokes}, 1}} \right)\cdot \bm{\hat{\phi}}
    \label{eq:bubblestokeslet}
\end{align}
with no modifications to $v_r$ and $v_z$ at leading order.
As in the case of the sphere, the higher order $1/\sphr^5$ term is necessary to satisfy Eq.~\ref{eq:vecpoisson} and the boundary condition. By equating the far-field flow and the Stokeslet solution, we find $F_z = 5\pi aU \mu$, which corresponds to the Stokesian drag on a bubble in a standard fluid; that is, the drag force is again unaffected at first order in odd viscosity. Note that the general form of the drag on a spherical bubble in a standard fluid is given by $F_z = 4\pi a U \mu_{\text{out}} \frac{\mu_{\text{out}} + \frac{3}{2}\mu_{\text{in}}}{\mu_{\text{out}} + \mu_{\text{in}}}$, where $\mu_{\text{out}}$ and $\mu_{\text{in}}$ are the even shear viscosities outside and inside the bubble, respectively. In the case we are considering, $\mu_{\text{in}} = \mu_{\text{out}}$, so $F_z$ reduces to the expression above \citep{batchelor2000introduction}.

Rewriting Eq.~\ref{eq:bubblestokeslet} more explicitly, we have
\begin{align}
    v_{\phi, 1}^{\text{out}} (\sphr, \theta) =\frac{U}{896}\left[  g^1_{\text{out}}(\theta) \frac{a}{\sphr} + g^3_{\text{out}}(\theta) \left(\frac{a}{\sphr} \right)^3 + g^5_{\text{out}}(\theta) \left(\frac{a}{\sphr} \right)^5 \right] \sin{2\theta} \label{eq:bubbleout}
\end{align}
where
\begin{align}
    g^1_{\text{out}}(\theta)=&\; 280 \epsilon_R^o - (175 + 210 \cos{2\theta}) \epsilon_1  - (70 + 105\cos{2\theta}) \epsilon_2 +\order{\epsilon^2} \\
    g^3_{\text{out}}(\theta)=& -168 \epsilon_R^o + ( 126 + 210 \cos{2\theta}) \epsilon_1 + (84 + 420\cos{2\theta}) \epsilon_2+\order{\epsilon^2} \\
    g^5_{\text{out}}(\theta)=&   -(15 + 105 \cos{2\theta}) \epsilon_1 - (30 + 210\cos{2\theta}) \epsilon_2 +\order{\epsilon^2}
\end{align}

Next, we consider the flow inside the bubble. In this case, $\Delta_{\alpha} \bm{v}_{\bm{0}}^{\text{in}} = 0$, so Eq.~\ref{eq:stokes} reduces to the vector Laplace equation,
\begin{align}
    \Delta \bm{v}_{\bm{1}}^{\text{in}} = 0
    \label{eq:veclaplace}
\end{align}
with the boundary condition $\bm{v}_{\bm{1}}^{\text{out}}(a, \theta) = \bm{v}_{\bm{1}}^{\text{in}}(a, \theta)$. The solution to this Dirichlet problem involves the Dirichlet Green function used in the sphere computation \citep{jackson1999classical} (see details in Appendix~\ref{app:bubble}). Solving for $\bm{v}_{\bm{1}}^{\text{in}}$, we find

\begin{align}
    v_{\phi, 1}^{\text{in}} (\sphr, \theta) = \frac{U}{56}\left[  g^1_{\text{in}}(\theta) \left(\frac{\sphr}{a}\right)^2 \right] \sin{2\theta} \label{eq:bubblein}
\end{align}
where
\begin{align}
    g^1_{\text{in}}(\theta)=& 7\epsilon_R^o - 4\epsilon_1  -  \epsilon_2 +\order{\epsilon^2}
\end{align}
again with no modifications to $v_r$ and $v_z$ at leading order. We plot the velocity fields outside and inside the bubble in Fig.~\ref{fig:spherebubble}H-J. 

\subsection{Effect of the hydrostatic torque}
\label{sec:rotsphere}

As we have mentioned in Section~\ref{nonsymmetric_stress_tensor}, parity-violating fluids, such as fluids made of spinning particles, can exhibit a hydrostatic torque $\sigma^\text{h}_{ij} = -\epsilon_{ijz} \tau_z$ and an hydrostatic shear stress density in their hydrostatic stress. 
Let us illustrate the effect of the hydrostatic torque on a finite sphere. We assume a no-slip boundary condition at the surface of the sphere. In contrast with the situation of Sec.~\ref{sec:sphere}, where the velocity at the surface of the sphere was assumed to vanish, here this velocity is determined by the balance between the torques due to the viscous stress and to the hydrostatic stress.
Note that other boundary conditions could be appropriate, depending on the microscopic interactions between the constituents of the fluid and the boundary.

In the absence of  odd viscosity, the hydrostatic torque leads to a total torque $T_z = \frac83 \pi a^3 \tau_z $ on the sphere (see Eq.~\ref{eq:net_torque}). Hence, the sphere rotates at a steady angular velocity $\Omega = \frac{T_z}{8 \pi a^3 \mu} =  \frac{\tau_z}{3 \mu}$ (see \cite{hobbie2007intermediate}) and introduces an additional azimuthal component
\begin{equation}
    v_\phi (\sphr, \theta)  = \frac{\Omega  a^3 \sin \theta}{\sphr^2}
\end{equation} 
to the flow. This expression is valid for all $\sphr$ (both in near and far field), as it satisfies the boundary condition. This azimuthal flow is even in $z$,  unlike the flow due to the parity-violating viscosities, which is odd in $z$.

When odd viscosity is present, we can still compute the far-field flow using the perturbative Oseen tensor (computed in Eq.~\eqref{perturbative_oseen_tensor} of Appendix~\ref{app:stokeslet}).
The far-field flow is given by $v_i =  A_{i z} T_z$ using the rotlet $A_{i \ell}$ (see Sec.~\ref{higher_order_responses}), and we find 
\begin{equation}
    \bm{v}
     = \frac{T_z}{8\pi\mu} \frac{\sin{\theta}}{\sphr^2}\bm{\hat{\phi}}
     +
     \epsilon \frac{T_z}{16\pi\mu}\frac{1 + 3\cos{2\theta}}{\sphr^2} \bm{\hat{\sphr}}
\end{equation}
in which $\epsilon \equiv \eta^o/\mu \ll 1$ and $\eta_2^o = - \eta_1^o /2 \equiv \eta^o$.
Again, we find an azimuthal flow even in $z$ as a consequence of the hydrostatic torque.
In addition, we see that the presence of odd viscosity combined with a hydrostatic torque generates flow in the radial direction that is absent in the zeroth order case.

\vspace{-2mm}
\section{Sedimentation in a parity-violating fluid}
\label{section_sedimentation}

\subsection{Few particles: mechanisms}
\label{sec:sediment}

\begin{figure}
 \begin{center}
 \leavevmode
\includegraphics[width=135mm]{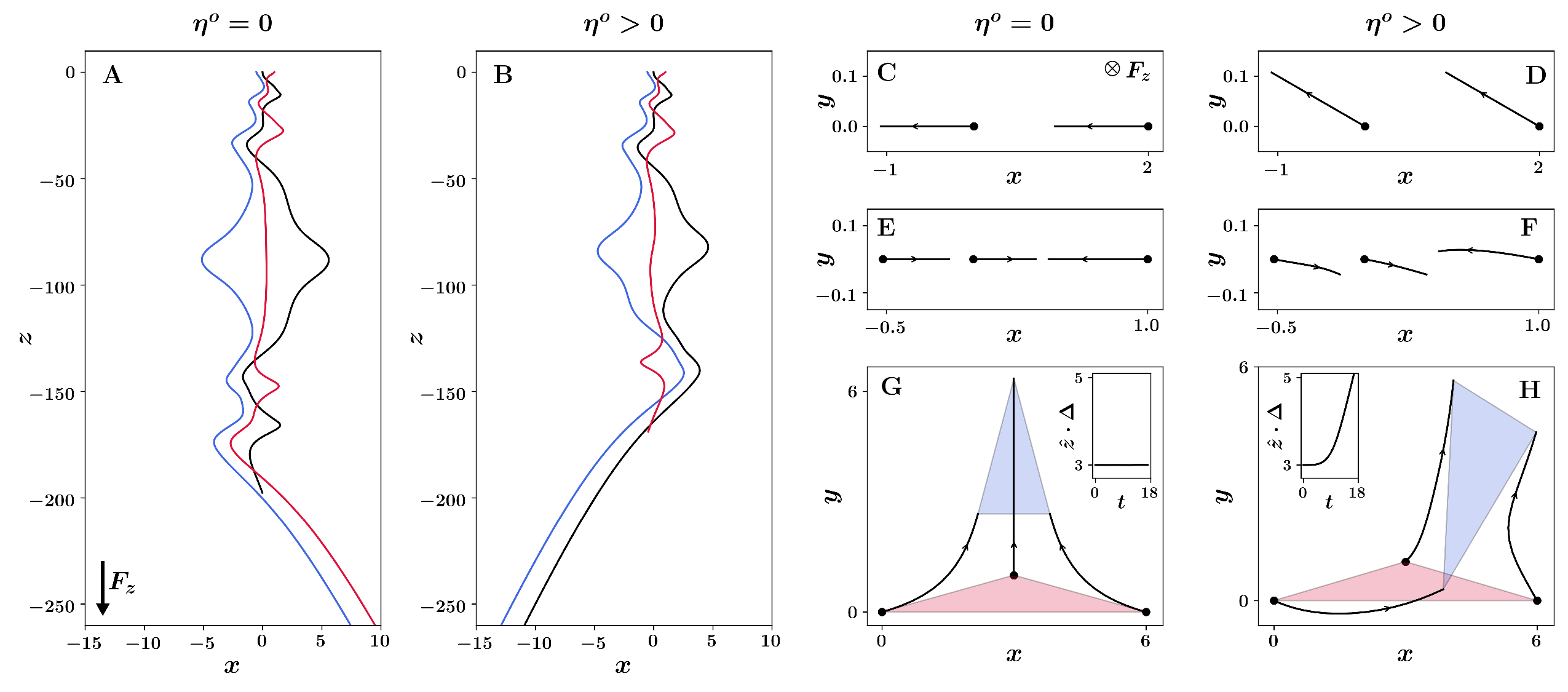} 
\caption{Few particle sedimentation in an odd viscous fluid. A-B. The trajectories of three particles, initially positioned along a horizontal line, without and with odd viscosity, respectively. Although the change to the trajectories is small, the addition of $\eta^o$ is sufficient to qualitatively change the long-time dynamics of the system, as the red particle gets left behind in panel B, while the black particle is lost in panel A. C-D. The trajectories of two particles initialized at different heights without and with odd viscosity, respectively. Rather than moving along the line connecting the two particles (panel C), the trajectories are rotated by an angle in the presence of $\eta^o$ (panel D). E-F. Three particle trajectories without and with odd viscosity, respectively. In panel E, the three particles remain in the same vertical plane as they sediment, but the addition of $\eta^o$ in panel F allows the particles to follow three-dimensional trajectories. G-H. The projected triangle area of a three particle system remains constant with time without odd viscosity (panel G), but can change with the addition of $\eta^o$ (panel F). Panels B, D, and F are computed with $\eta^o = 0.1$, panel H with $\eta^o = 1$.}
\label{fig:sedfew}
\end{center}
\end{figure}

 We now examine the role of parity-violating viscosities and of the corresponding azimuthal flows on the problem of sedimentation, in which particles driven by an external field (e.g. gravity) interact hydrodynamically at low Reynolds number. We assume the particles to be small identical spheres without inertia, which sediment under gravity and are advected by the flow due to the other particles.

In a standard isotropic fluid, an isolated sedimenting particle experiences a Stokesian drag and thus sinks at the velocity $U = F_z/(6\pi \mu a)$. In a co-moving reference frame, the velocity field generated by a single sedimenting sphere $\beta$ is simply given by the Stokes flow past a sphere (Eq.~\ref{eq:appnormsphere}). 
In the dilute limit, we can neglect the near field terms that fall off faster than $1/\sphr$ \citep{Happel1983low}. In particular, we neglect the higher order $1/\sphr^2$ velocity field contribution associated with particle rotation (see Sec.~\ref{sec:rotsphere}), which may occur in a fluid with hydrostatic torques.
As a result, the velocity field generated by each particle simply reduces to the Stokeslet (Eq.~\ref{eq:appstokesnorm}).
If all the sedimenting particles experience the same force $\bm{f} = -\bm{\hat{z}}F_z$, the equation of motion in the co-moving frame for particle $\alpha$ becomes~\citep{hocking1964behaviour,Guazzelli2009}
\begin{equation}
    \frac{d\bm{x}^{\alpha}}{dt} = \sum_{
    \alpha \neq \beta} G(\bm{x}^{\alpha} - \bm{x}^{\beta})\bm{f}
    \label{eq:sediment}
\end{equation}
where $G(\bm{x})$ is the Green function of the Stokes equation (Oseen tensor) from Eq.~\ref{eq:realspace}, $\alpha$ and $\beta$ are particle indices, and $\bm{x}^\alpha$ is the position of particle $\alpha$ in the co-moving reference frame.

In a parity-violating fluid,  we replace the standard Stokeslet field on the right hand side of Eq.~\ref{eq:sediment} with the odd viscous Stokeslet from Eqs.~\ref{eq:str}-\ref{eq:stz}. For simplicity, here we will consider  $\eta_2^o = - \eta_1^o /2 \equiv \eta^o$ {and $\mu_1 = \mu_2 = \mu_3 \equiv \mu$ (all other viscosities in Eq.~\ref{eq:planarchiral} are set to zero)}. 
Since each of the sedimenting particles experiences an identical vertical force, we can move into their comoving reference frame. 
We then numerically integrate Eq.~\ref{eq:sediment} over time with a standard fourth order Runge-Kutta algorithm to obtain the trajectories of the particles.

Figure~\ref{fig:sedfew}A-B shows the trajectories of three particles in the $x$-$z$ plane with $\eta^o=0$ and $\eta^o >0$.  
While the sedimentation of as few as three Stokeslets in a standard isotropic fluid is already chaotic~\citep{hocking1964behaviour, janosi1997chaotic}, the parity violating flow introduces simple and well-defined modifications to the trajectories. 
For example, the trajectories of two particles interacting through the standard Stokeslet are confined to the vertical plane (Fig.~\ref{fig:sedfew}C) containing initial positions. In the presence of odd viscosity, the particle trajectories are deflected out of this plane due to the azimuthal flow present in the odd Stokeslet (Fig.~\ref{fig:sedfew}D). 
Similarly, the dynamics of a three particle system initialized along a horizontal line is constrained to the initial vertical plane, as shown in Fig.~\ref{fig:sedfew}E \citep{hocking1964behaviour}. The azimuthal flow in the odd Stokeslet allows the trajectories to escape out of this plane and follow three-dimensional trajectories (Fig.~\ref{fig:sedfew}F).
More generally, the area, ${\bm \Delta} \cdot {\bm \hat z}$, of the projection on the $x$-$y$ plane of the triangle formed by three sedimenting particles is constant in time for standard fluids~\citep{hocking1964behaviour}, even though its shape will generally change (Fig.~\ref{fig:sedfew}G). This conservation law is broken when azimuthal flow is present, as illustrated in Fig.~\ref{fig:sedfew}H. In the figure, we have used a high $\eta^o = 1$ for ease of visualization, but the effect is also present in the perturbative regime.

\subsection{Sedimentation of clouds}

\begin{figure}
 \begin{center}
 \leavevmode
\includegraphics[width=135mm]{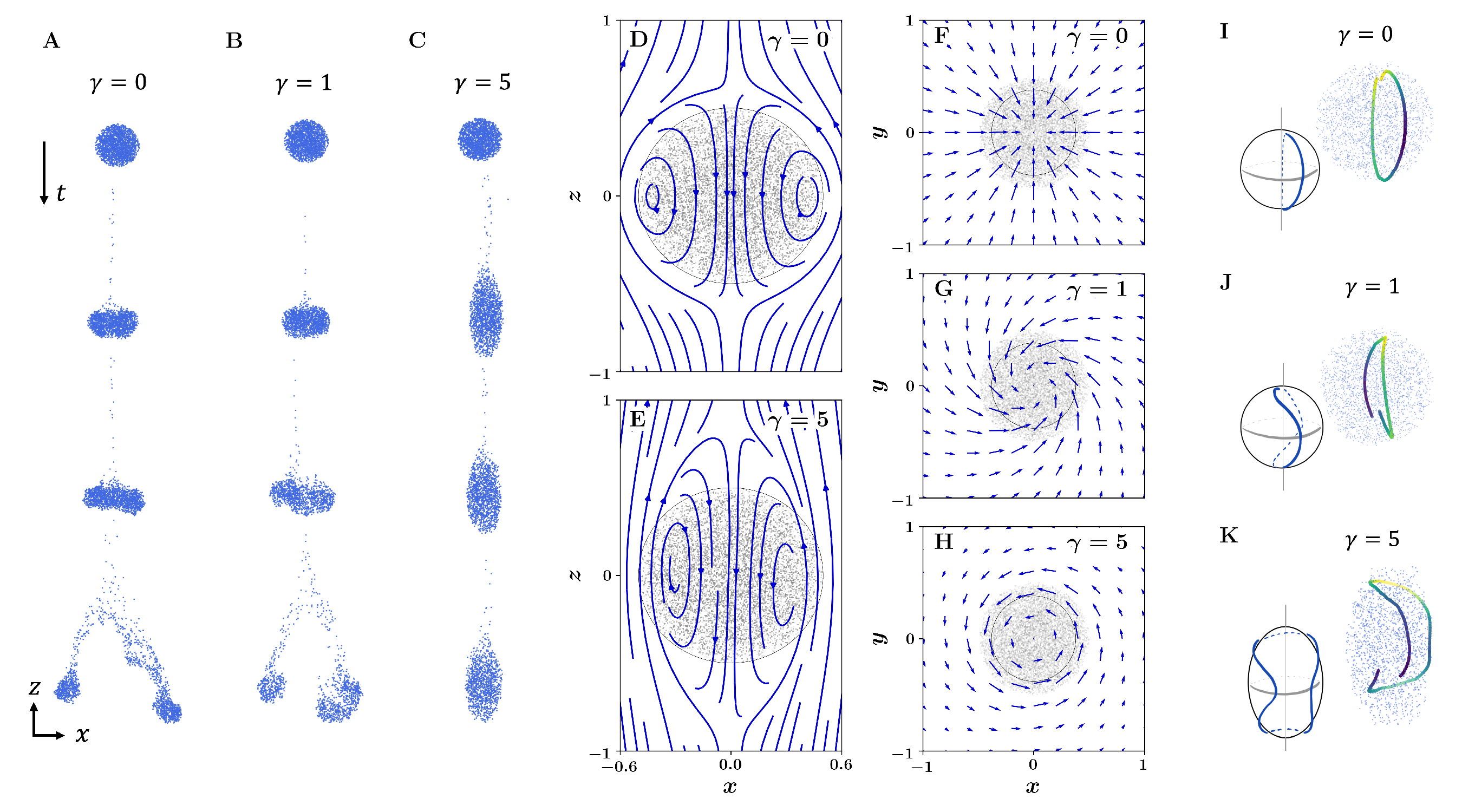} 
\caption{Sedimentation of a cloud in an odd viscous fluid. A-C. Snapshots of the falling cloud from simulations with $N = 2000$ particles for different values of $\gamma = \eta^o/\mu$. In the absence of odd viscosity, the initially spherical cloud deforms into a torus, and subsequently breaks apart into smaller clouds. As the odd viscosity is increased, the break-up event only occurs in a fraction of the runs, and for even higher values of $\gamma$, the cloud no longer forms a torus, instead deforming into an ellipsoid. D-E. Streamlines of the fluid flow in the $x$-$z$ plane with $y=0$, computed at $t=0$ in the instantaneous reference frame of the cloud. When $\gamma = 0$, the flow field corresponds to Hill's spherical vortex. In contrast, for a large odd viscosity, the initially spherical cloud immediately deforms to an ellipsoid due to the stretched vortices. F-H. The velocity field in the $x$-$y$ plane with $z = 0.33$, computed at $t = 0$, for various values of $\gamma$. As the odd viscosity is increased, the radial component of the velocity decreases while the azimuthal component increases. I-K. Sample particle trajectories for varying $\gamma$, with supplementary schematics to highlight the main features. The color map indicates distance from the viewer, with dark blue closest.}
\label{fig:sedcloud}
\end{center}
\end{figure}

We now consider a cloud composed of many sedimenting particles. We start with $N = 2000$ particles uniformly distributed within a spherical volume of radius $a = 0.5$, and evolve the system by integrating Eq.~\ref{eq:sediment}. 
Figure~\ref{fig:sedcloud}A-C shows snapshots of the evolution for different values of odd viscosity (see also Supplementary Movie 2, described in Appendix~\ref{app:movies}).

The case of a standard fluid (Fig.~\ref{fig:sedcloud}A) was analyzed theoretically and experimentally by \cite{batchelor1972sedimentation, nitsche1997break, ekiel2006spherical, metzger2007falling} (see also references therein). In this case, the cloud develops a vertical tail of particles that are lost from the outside layer of the cloud. Then, the circulating motion of the flow inside the cloud (Fig.~\ref{fig:sedcloud}D) depletes the number of particles along the central vertical axis of the cloud, leading to the formation of a torus. If the initial cloud is sufficiently large, the torus undergoes a breakup event into smaller clouds. 
As shown in Fig.~\ref{fig:sedcloud}D, the streamlines of the fluid velocity field due to the particles (or, equivalently, the trajectories of the particles themselves, since inertia is neglected) initially coincide with the flow past a bubble from Section \ref{sec:bubble} \citep{ekiel2006spherical,kojima1984formation, pozrikidis1990instability, shimokawa2016breakup}. 

We now analyze the effect of a non-zero odd viscosity.
In the regime of small $\gamma = \eta^o/\mu$, the streamlines inside the cloud still agree with the velocity field in a bubble from Eq.~\ref{eq:bubblein}. Particles develop a small in-plane tangential velocity component that points in opposite directions below and above the equator. This perturbative modification due to odd viscosity does not yet affect the qualitative features of the cloud dynamics (i.e. the formation of a torus and its breakup). As odd viscosity is increased further, the strength of the azimuthal flow increases in comparison with the radial component, and qualitative features of the evolution begin to change. At $\gamma = 1$, for instance, the break-up event does not always happen. At $\gamma = 5$, we find that the cloud no longer deforms into a torus. Instead, the cloud adopts an ellipsoidal shape which persists until all particles have leaked into the trailing tail (Fig.~\ref{fig:sedcloud}C). The formation of the ellipsoid is visible from the initial flow within the cloud; unlike the Hill's spherical vortex shown in Fig.~\ref{fig:sedcloud}D, the initial streamlines in the high odd viscosity case form a stretched vortex flow (Fig.~\ref{fig:sedcloud}E). The radial and azimuthal flows, as seen from above, are shown in Fig.~\ref{fig:sedcloud}F-H.  

In all cases, the particles within the cloud follow approximately closed trajectories (Fig.~\ref{fig:sedcloud}I-K). In the absence of odd viscosity, the closed loop trajectories are angled radially inward to the central axis of the cloud (Fig.~\ref{fig:sedcloud}I). When odd viscosity is nonzero, the loop deforms due to the azimuthal flow. Since the azimuthal component changes sign above and below the equator of the cloud, the particles change rotation direction along the trajectories, now creating curved closed loops (see Fig.~\ref{fig:sedcloud}J). For values of $\gamma$ that correspond to ellipsoidal clouds, the trajectories wrap around layers of the cloud with little radial motion, again rotating in opposite directions above and below the equator (Fig.~\ref{fig:sedcloud}H). 

\medskip

\section{Conclusion} 

In this article, we have explored the effects of parity violation on the viscous response of a fluid in three dimensions. The broken mirror symmetry gives rise to azimuthal flows even when the external forcing is aligned with the axis of azimuthal symmetry. The changes in a single Stokeslet lead to qualitative changes in the sedimentation of both few and many particles. 
The situations we have analyzed theoretically and numerically are within experimental reach. In the context of soft matter, this could be done in multiple-scale colloids: a colloidal suspension of rotating particles can produce effective fluids with parity-violating viscosities (as it was already demonstrated in 2D ~\citep{soni2019odd}), while larger particles can act as colloidal particles for the effective fluid. 
In these systems, the presence of parity-violating coefficients could also affect hydrodynamic instabilities such as the fingering instabilities observed in colloidal rollers in suspension~\citep{Wysocki2009,Driscoll2016}.
In the context of hard condensed matter, recent experimental and theoretical works~\citep{hoyos2012hall,Levitov2016,Holder2019} have focused on the hydrodynamic behavior of electrons in solids. There, sizeable parity-violating viscosities can occur and have been observed when the sample is under a magnetic field~\citep{Berdyugin2019measuring} and Stokes flow can be realized by introducing holes in the sample~\citep{Gusev2020}.

\medskip
\textbf{Declaration of Interests.} The authors report no conflict of interest.

\appendix

\vspace{-5mm}

\section{Coordinate systems and Fourier transform conventions}
\label{app_coordinate_systems}

In this Appendix, we give explicit expressions for the coordinate systems used in this work (see also the schematic in Figure~\ref{fig:stokesletexact}A).
Writing Cartesian coordinates $(x, y, z)$ in terms of the cylindrical coordinates $(r, \phi, z)$, we have
\begin{subequations}
\begin{align}
    x &= r \cos{\phi}\\
    y &= r \sin{\phi}\\
    z &= z
\end{align}
\end{subequations}
In terms of the spherical coordinates $(\sphr, \phi, \theta)$, we have
\begin{subequations}
\begin{align}
    x &= \sphr \, \sin{\theta} \cos{\phi}\\
    y &= \sphr \, \sin{\theta} \sin{\phi}\\
    z &= \sphr \, \cos{\theta}
\end{align}
\end{subequations}
Finally, we define the Fourier transform of a function $\phi$ as follows
\begin{align}
    \phi(\bm{x}) &= \frac{1}{(2\pi)^3}\int \dd^3 q \phi(\bm{q}) e^{i \bm{q}\cdot \bm{x}} \\
    \phi(\bm{q}) &= \int \dd^3 q \phi(\bm{x}) e^{-i \bm{q}\cdot \bm{x}} 
\end{align}

\vspace{-5mm}
\section{Symmetries}\label{app_symmetries}

The action of an isometry $R \in O(3)$ on the stress and strain rate tensors is
\begin{equation}
    \label{transformation_rank_two_tensors}
    \sigma_{i j} \mapsto R_{i i'} R_{j j'} \sigma_{i' j'}
    \qquad
    \text{and}
    \qquad
    \dot{e}_{i j} \mapsto R_{i i'} R_{j j'} \dot{e}_{i' j'}
\end{equation}
which can be written as $\sigma \mapsto R \sigma R^T$ and $\dot{e} \mapsto R \dot{e} R^T$ in matrix notation.

The basis of tensors $\tau^A$ intorduced in  Eq.~\ref{irrep_components_from_rank_two_tensors} arises from the decomposition in irreducible representations (irrep) of the action Eq.~\ref{transformation_rank_two_tensors} of $O(3)$ on rank two tensors. 
Here, the (proper or improper) rotation $R \in O(3)$ correspond to the vector representation $D_{1}^{-}$ of $O(3)$, in which $D_{\ell}^{+(-)}$ are the positive (negative) irreducible representations of $O(3)$ with dimension $(2 \ell + 1)$ [i.e. with angular momentum $\ell$] and parity $\pm$ (see e.g. \cite{Altmann2013,Miller1973}).
Hence, the representation given by Eq.~\ref{transformation_rank_two_tensors} is
\begin{equation}
    D_{1}^{-} \otimes D_{1}^{-} \simeq D_{0}^{+} \oplus D_{1}^{+} \oplus D_{2}^{+}
\end{equation}
The basis tensor $C$ corresponds to the 1D irrep $D_{0}^{+}$, the basis tensors $R^k$ to the 3D irrep $D_{1}^{+}$, and the basis tensors $S^k$ to the 5D irrep $D_{2}^{+}$.

In terms of the components of the decomposition of the stress or strain rate on this basis (defined in Eq.~\ref{irrep_components_from_rank_two_tensors}), the action Eq.~\ref{transformation_rank_two_tensors} reads
\begin{equation}
    \sigma^A \mapsto \mathcal{R}^{A B} \sigma^B
    \qquad
    \text{and}
    \qquad
    \dot{e}^A \mapsto \mathcal{R}^{A B} \dot{e}^B
\end{equation}
in which
\begin{equation}
    \label{action_rotation_on_rank_two_tensors}
    \mathcal{R}^{A B} = \frac{1}{2} \, \tau^{A}_{i j} R_{i i'} R_{j j'} \tau^B_{i' j'}
\end{equation}
is an orthogonal matrix. 
Under this transformation, the viscosity matrix transforms as
\begin{equation}
    \eta^{A B} \mapsto \mathcal{R}^{A A'} \mathcal{R}^{B B'} \eta^{A' B'}.
\end{equation}

For example, consider a reflection over the $y$-axis, whose action on $\mathbb{R}^3$ is given by the matrix 
\begin{equation}
    P_y = 
    \begin{bmatrix}
    1 & 0 & 0 \\
    0 & -1 & 0 \\
    0 & 0 & 1
    \end{bmatrix}
\end{equation}
We can then compute the action on the basis tensors. For instance, shear two transforms as follows, 
\begin{align}
    P_y S_2 P_y^T = 
    \begin{bmatrix}
    1 & 0 & 0 \\
    0 & -1 & 0 \\
    0 & 0 & 1
    \end{bmatrix}
    \begin{bmatrix}
    0 & 1 & 0 \\
    1 & 0 & 0 \\
    0 & 0 & 0
    \end{bmatrix}
    \begin{bmatrix}
    1 & 0 & 0 \\
    0 & -1 & 0 \\
    0 & 0 & 1
    \end{bmatrix}
    = 
    \begin{bmatrix}
    0 & -1 & 0 \\
    -1 & 0 & 0 \\
    0 & 0 & 0
    \end{bmatrix}
    =-S_2
\end{align}

Considering all of the basis matrices in this way allows us to construct the matrix
\begin{equation}
\mathcal{P}_y =
    \begin{bmatrix}
    -1 & 0 & 0 & 0 & 0 & 0 & 0 & 0 & 0\\
    0 & 1 & 0 & 0 & 0 & 0 & 0 & 0 & 0\\
    0 & 0 & -1 & 0 & 0 & 0 & 0 & 0 & 0\\
    0 & 0 & 0 & 1 & 0 & 0 & 0 & 0 & 0\\
    0 & 0 & 0 & 0 & 1 & 0 & 0 & 0 & 0\\
    0 & 0 & 0 & 0 & 0 & -1 & 0 & 0 & 0\\
    0 & 0 & 0 & 0 & 0 & 0 & 1 & 0 & 0\\
    0 & 0 & 0 & 0 & 0 & 0 & 0 & -1 & 0\\
    0 & 0 & 0 & 0 & 0 & 0 & 0 & 0 & 1\\
    \end{bmatrix}
\end{equation}
given by Eq.~\ref{action_rotation_on_rank_two_tensors}, that describes how the stress and strain rate transform under reflection across the $y$-axis.
Consequently, the viscosity matrix transforms under this reflection as 
\begin{equation}
    \eta \mapsto \eta' = \mathcal{P}_y \eta \mathcal{P}_y^T.
\end{equation}

\begin{figure}
    \centering
    \includegraphics{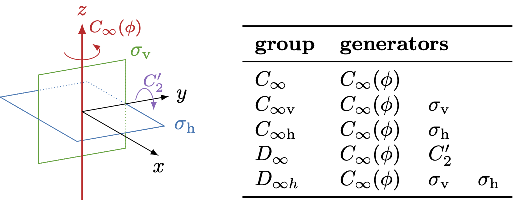}
    \caption{\label{symmetry_ops} Symmetry operations used to define the groups with axial symmetry.
    Here, $C_{\infty}(\phi)$ are rotations about the $z$ axis (by an angle $\phi$ ; in red), $\sigma_{\text{v}}$ is a mirror reflection about a plane containing the $z$ axis (in green), $\sigma_{\text{h}}$ is a mirror reflection about a plane orthogonal to the $z$ axis (in blue), and $C_2'$ is a 2-fold ({180}) rotation about any axis orthogonal to the $z$ axis (purple).
    We also give generators of the five infinite axial groups.
    We follow IUPAC recommendations~\citep{Schutte1997} for the point groups and symmetry operations, given with Schoenflies notations.
    See \cite{Shubnikov1988} and \cite{ITA} (in particular Table~\S~10.1.4.2 p.~799 and Fig.~\S~10.1.4.3 p.~803) for more details, including the correspondence with Hermann-Mauguin notations.
    }
\end{figure}

For each symmetry group in Fig.~\ref{fig:descent_symmetry}, the allowed viscosity coefficients are derived by requiring that $\eta_{ijk\ell}$ be invariant under all the corresponding generators listed in the table in Fig.~\ref{symmetry_ops}.
The generators in the table of Figure~\ref{symmetry_ops} can be represented explicitly as matrices
\begin{align}
    C_\infty(\phi) =& \begin{bmatrix}
                    \cos \phi  & \sin \phi & 0 \\
                    -\sin \phi & \cos \phi & 0 \\
                    0 & 0 & 1
                  \end{bmatrix}
    &
    \sigma_v &= \begin{bmatrix}
                    -1  & 0 & 0 \\
                    0 & 1 & 0 \\
                    0 & 0 & 1
        \end{bmatrix}
    \\ 
    \sigma_h &= \begin{bmatrix}
        1 & 0 & 0 \\
        0 & 1 & 0 \\
        0 & 0 & -1 
        \end{bmatrix}
    &
    C_2' &= \begin{bmatrix}
        -1 & 0 & 0  \\
        0 & 1 & 0 \\
        0 & 0 & -1
    \end{bmatrix}
\end{align}
acting on $\mathbb{R}^3$ (following the vector representation $D_1^{-}$), in which the $C_2'$ axis has been chosen so that $C_2' = \sigma_h \sigma_v$. 
Notice that the viscosity tensor is automatically invariant under $ -\one = \operatorname{diag}(-1,-1,-1)$ since four copies of the $- \one$ cancel in Eq.~(\ref{viscosity_tensor_transformation}). Notice that $\sigma_h = - \one \cdot C_\infty(\frac\pi2)$. Hence, any cylindrical symmetric viscosity tensor is invariant under $\sigma_h$ regardless of whether $\sigma_h$ is an element of the underlying symmetry group of the fluid. Thus, the effective symmetry group of the viscosity tensor is generally larger than the symmetry group of the fluid, and it can be obtained by simply adding $\sigma_h$ to the list of generators of the symmetry group of the underlying fluid. 

\section{Energy dissipation}

\subsection{Dissipated power}

\label{app:energy}
Let us start from the Stokes equation \eqref{stokes_intro}
\begin{equation}
    \rho \partial_t v_i = \partial_j \sigma_{i j} + f_i
\end{equation}
The total stress tensor is split into two pieces
\begin{equation}
    \sigma_{ij} = \sigma_{ij}^\text{h}+\sigma_{ij}^\text{vis}
\end{equation}
where $\sigma_{ij}^\text{h}$ are the ``hydrostatic" stresses present even when there are no velocity gradients while the viscous stresses $\sigma_{ij}^\text{vis} = \eta_{ijk\ell} \partial_\ell v_k$ arise as a linear response to velocity gradients.
(This decomposition is distinct from the geometric decomposition of the stress tensor $\sigma_{i j} = \pi \delta_{i j} + s_{i j}$ into a volumetric stress $\pi \equiv \sigma_{i i}/d$ ($d$ is the space dimension) and a deviatoric (i.e. traceless) stress $s_{i j} \equiv \sigma_{i j} - \pi \delta_{i j}$.) 

After multiplying by $v_i$ and integrating the result over a volume $\VV$, we obtain
\begin{equation}
    \int_{\VV} \rho \, \partial_t \left( \frac{v^2}{2} \right) \dd^3 x 
    = \int_{\VV} v_i (\partial_j \sigma_{i j}) \dd V + \int_{\VV} v_i f_i \dd^3 x.
\end{equation}
After an integration by parts (ignoring boundary terms for simplicity), we obtain
\begin{align}
    \int_{\VV} \rho \, \partial_t \left( \frac{v^2}{2} \right) \dd^3 x 
    =& - \int_{\VV} (\partial_j v_i) \sigma_{i j} \dd^3 x + \int_{\VV} v_i f_i \dd^3 x. \\
    =&- \int_{\VV} (\partial_j v_i) \sigma_{i j}^\text{h} \dd^3 x - \int_{\VV} (\partial_j v_i) \sigma_{i j}^\text{vis} \dd^3 x + \int_{\VV} v_i f_i \dd^3 x.
\end{align}
Here, $(\partial_j v_i) \sigma_{ij}^\text{h}$ is the rate of change of stored energy in the fluid element.
This allows us to identify 
\begin{equation}
    \dot{w} \equiv (\partial_j v_i) \sigma_{i j}^\text{vis} 
\end{equation}
as the local rate of energy dissipation in the fluid.
Finally, we obtain
\begin{equation}
\dot{w}
= (\partial_j v_i) \sigma_{i j}^{\text{vis}} = \eta_{i j k \ell} \, (\partial_j v_i) \, (\partial_\ell v_k) = \eta_{i j k \ell}^{\text{e}} \, (\partial_j v_i) \, (\partial_\ell v_k) \label{eq:diss}
\end{equation}
which can be interpreted as the rate of viscous dissipation in the fluid, and in which only the symmetric part of the viscosity tensor contributes, by symmetry of the expression.

In terms of the viscosity matrix defined in Eq.~\ref{viscosity_matrix_def}, the dissipated power reads
\begin{equation}
    \label{dissipated_power_matrix}
    \dot{w} = \frac{1}{2} \, \eta^{A B} \dot{e}^A \dot{e}^B = \frac{1}{2} \,  [\eta^{\text{e}}]^{A B}  \dot{e}^A \dot{e}^B.
\end{equation}
in which $\eta^{\text{e}} = (\eta + \eta^{\text{T}})/2$ is the symmetric part of the viscosity matrix, and $A$, $B$ label its components.

\subsection{Virtual power and Lorentz reciprocity}

We now note that Lorentz reciprocity~\citep{Happel2012,masoud2019reciprocal}, which was defined in the main text in terms of Green functions, can be interpreted from an energetic viewpoint.
To do so, consider two a priori unrelated incompressible velocity fields $\vec{v}$ and $\vec{v'}$ satisfying the Stokes equation, and the corresponding stress tensors $\sigma_{ij} = - P \delta_{i j} + \eta_{i j k \ell} \partial_\ell v_k$ and $\sigma_{ij}' = - P' \delta_{i j} + \eta_{i j k \ell} \partial_\ell v_k'$.
We consider the quadratic form
\begin{equation}
    \dot{W}[\vec{v}, \vec{v'}] = \sigma_{ij} \, \partial_j v_i' = \eta_{i j k \ell} \, \partial_\ell v_k \, \partial_j v_i'
\end{equation}
in which we used $\partial_k v_k' = 0$.
The quantity $\dot{W}[\vec{v}, \vec{v'}]$ can be seen as the virtual power exerted by the stress tensor $\sigma_{i j}$ in the velocity field $\vec{v'}$, and $\dot{W}[\vec{v}, \vec{v}]$ reduces to the local power $\dot w$ dissipated  in the fluid as given by Eq.~\ref{eq:diss}.
Permuting the arguments, we get
\begin{equation}
    \dot{W}[\vec{v'}, \vec{v}] = \sigma_{ij}' \, \partial_j v_i = \eta_{i j k \ell} \, \partial_\ell v_k' \, \partial_j v_i 
    = \eta_{k \ell i j} \, \partial_\ell v_k \, \partial_j v_i'.
\end{equation}
Hence, the reciprocity theorem $\dot{W}[\vec{v}, \vec{v'}] = \dot{W}[\vec{v'}, \vec{v}]$ is in general satisfied only when $\eta_{i j k \ell} = \eta_{k \ell i j}$, namely when the viscosity tensor is purely dissipative. 

\subsection{Positivity of the dissipated power}

Taking the Fourier transform of equation~(\ref{eq:diss}) yields:
\begin{align}
    \dot{w}({\bm q}, \omega) = -  v_i^\dagger({\bm q, \omega})  M_{ik}({\bm q}, \omega) v_k ({\bm q}, \omega).  \label{eq:r} 
\end{align}
In the derivation of the Green function in Eq.~\ref{eq:gen}, we assumed that $M$ is negative definite for all ${\bm q} \neq 0$. (This implies that $M$ is invertible at finite ${\bm q}$, as used in the derivation). On the one hand, from Eq.~\ref{eq:r}, we see that requiring  $M$ to be negative definite is equivalent to requiring the dissipation rate be strictly positive for all flows at finite ${\bm q}$. 
On the other hand, Eq.~\ref{eq:diss} and Eq.~\ref{dissipated_power_matrix} show that a necessary and sufficient condition for $\dot{w} > 0$ is that $\eta^\text{e}_{ijk\ell}$ is a positive definite linear map on the space of rank two tensors, or equivalently that the symmetric part of the viscosity matrix $\eta^{A B}$ is positive definite.

\section{Incompressible Stokes flow in two dimensions}
\label{app:2D}

In a two dimensional isotropic fluid in which $\eta_{ijk\ell}$ retains both its minor symmetries, odd viscosity is captured by a single coefficient:
\begin{align}
    \eta_{ijk\ell}^o = \frac{\eta^o}{2} (\epsilon_{ik} \delta_{j\ell} + \epsilon_{i\ell} \delta_{jk} + \epsilon_{jk} \delta_{i\ell} + \epsilon_{j\ell} \delta_{ik}) 
\end{align}
In this case, the odd viscosity enters the equations of motion for the velocity field as:
\begin{align}
    \rho D_t v_i = - \partial_i P + \qty(\xi+\frac13 \mu) \partial_i \partial_j v_j + \mu \Delta v_i + \eta^o\epsilon_{ij} \Delta  v_j 
\end{align}
Using $\epsilon_{ij} \Delta v_j
= - \partial_i[ \epsilon_{k \ell} \partial_k v_\ell]
$ for an incompressible fluid (for which $\partial_i v_i = 0$), we can rewrite the equations of motion for the velocity field as
\begin{align}
\rho D_t v_i =& -\partial_i \tilde P + \mu \Delta v_i \\
\partial_i v_i =& 0
\end{align}
in which $\tilde P = P + \eta^o \epsilon_{k \ell} \partial_k v_\ell$ is an effective pressure. Since odd viscosity drops out of the bulk equations of motion, it does not affect the flow of an incompressible, isotropic 2D fluid unless boundary conditions on the fluid are stated in terms of stresses, not velocities (e.g., at a free surface)~\citep{avron1998odd,banerjee2017odd,Abanov2019free}. For this reason, the Stokeslet flow is not modified in two dimensions. However, odd viscosity does modify the flow of incompressible anisotropic 2D fluids~\citep{Souslov2020anisotropic}. 

In three dimensions, it is not possible to absorb the odd shear viscosity terms in the pressure. The form of the odd terms in the Stokes equation is given in Appendix~\ref{app:oddop}; since these terms cannot be written as gradients of a scalar function, we expect that $\eta_1^o$ and $\eta_2^o$ can in fact lead to changes in the velocity field, in agreement with the results demonstrated in the main text.

\section{Modification to Stokes flow}
\label{app:oddop}

The steady Stokes equation for an incompressible fluid reads
\begin{equation}
    0 = - \partial_i P + \partial_j [\eta_{i j k \ell} \partial_\ell v_k] + f_i
    \quad
    \text{with}
    \quad
    \partial_i v_i = 0.
\end{equation}

Once the form of $\eta_{ijkl}$ is specified, we write the viscous term using shorthand vector notation to distinguish between the even and odd viscosity contributions. In this notation, the Stokes equation becomes
\begin{equation}
    0 = -\nabla P + \mu \Delta \bm{v} + \alpha \Delta_{\alpha} \bm{v}
\end{equation}
in which $\Delta_{\alpha}$ is the second-order differential operator associated to the viscosity $\alpha=\eta_1^o, \eta_2^o$, and $\eta_R^o $. 
In Cartesian coordinates, they are given by
\begin{align}
    \Delta_{\eta_1^o}\bm{v}
    =
    \begin{bmatrix}
        (\partial_x^2 + \partial_y^2)v_y\\
        \\
        -(\partial_x^2 + \partial_y^2)v_x\\
        \\
        0
    \end{bmatrix} \;
     \Delta_{\eta_2^o}\bm{v}
    =
    \begin{bmatrix}
         -\partial_z^2 v_y - \partial_y \partial_z v_z\\
        \\
         \partial_z^2 v_x + \partial_x \partial_z v_z\\
        \\
        \partial_z(\partial_y v_x - \partial_x v_y)
    \end{bmatrix} \;
    \Delta_{\eta_R^o}\bm{v}
    =
    \begin{bmatrix}
        \partial_z \omega_x\\
        \partial_z \omega_y\\
        -\partial_x \omega_x - \partial_y \omega_y
    \end{bmatrix}
\end{align}
where $\bm{\omega}$ is the vorticity.

Assuming that $\bm{v}$ has no dependence on $\phi$, the expressions of $\Delta_{\alpha}$ in cylindrical coordinates are
\begin{align}
    \Delta_{\eta_1^o}\bm{v} =
    \begin{bmatrix}
        \partial_r^2 v_{\phi} + \frac{\partial_r v_{\phi}}{r}-\frac{v_{\phi}}{r^2}\\
        \\
        -\partial_r^2 v_r - \frac{\partial_r v_r}{r} + \frac{v_r}{r^2}\\
        \\
        0
    \end{bmatrix}
    \;
   \Delta_{\eta_2^o}\bm{v} =
    \begin{bmatrix}
         - \partial_z^2 v_{\phi}\\
        \\
         \partial _r \partial_z v_z + \partial_z^2 v_r\\
        \\
        -\frac{\partial_z v_{\phi}}{r} - \partial_r \partial_z v_{\phi}
    \end{bmatrix}
    \;
    \Delta_{\eta_R^o}\bm{v} =
    \begin{bmatrix}
        \partial_z \omega_r\\
        \partial_z \omega_{\phi}\\
        -\frac{\omega_r}{r}-\partial_r \omega_r
    \end{bmatrix}
\end{align}

\section{Stokeslet: numerical solution}
\label{app:numerics}

\begin{figure*}
 \begin{center}
 \leavevmode
\includegraphics[width=100mm]{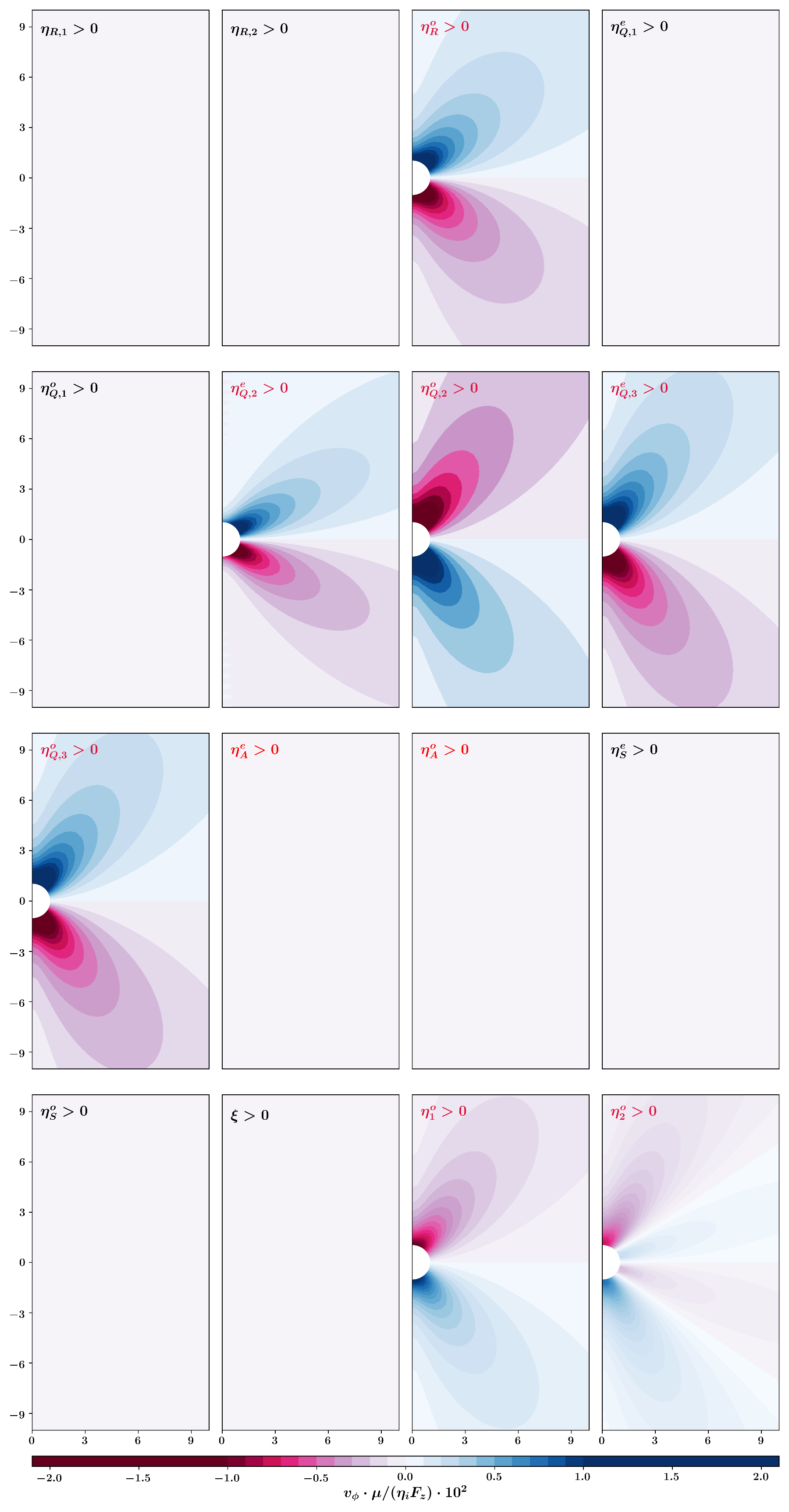}
\caption{The azimuthal component of the Stokeslet flow, computed numerically for all viscosity coefficients allowed by cylindrical symmetry. The parity-violating viscosities are labeled in red.
The azimuthal component is non-zero only for parity-violating viscosities.
The coefficients $\eta_A^{\text{e}/\text{o}}$ are parity-violating, but do not lead to an azimuthal flow, respectively because the flow is incompressible and because the corresponding term in the Navier-Stokes equation can be absorbed in pressure (see main text).
}

\label{fig:numerics}
\end{center}
\end{figure*}

In Fig.~\ref{fig:numerics}, we visualize the azimuthal component of the Stokeslet velocity field on the $r$-$z$ plane for all viscosity coefficients in Eq.~\ref{eq:planarchiral}. Each solution is computed numerically, as outlined in Section~\ref{sec:numerics}, in the presence of the shear viscosity $\mu$ and an additional viscosity, indicated in the text label on each panel of Fig.~\ref{fig:numerics}. Here, each such $\eta_i = 0.01 \mu$.

\begin{figure*}
 \begin{center}
 \leavevmode
\includegraphics[width=120mm]{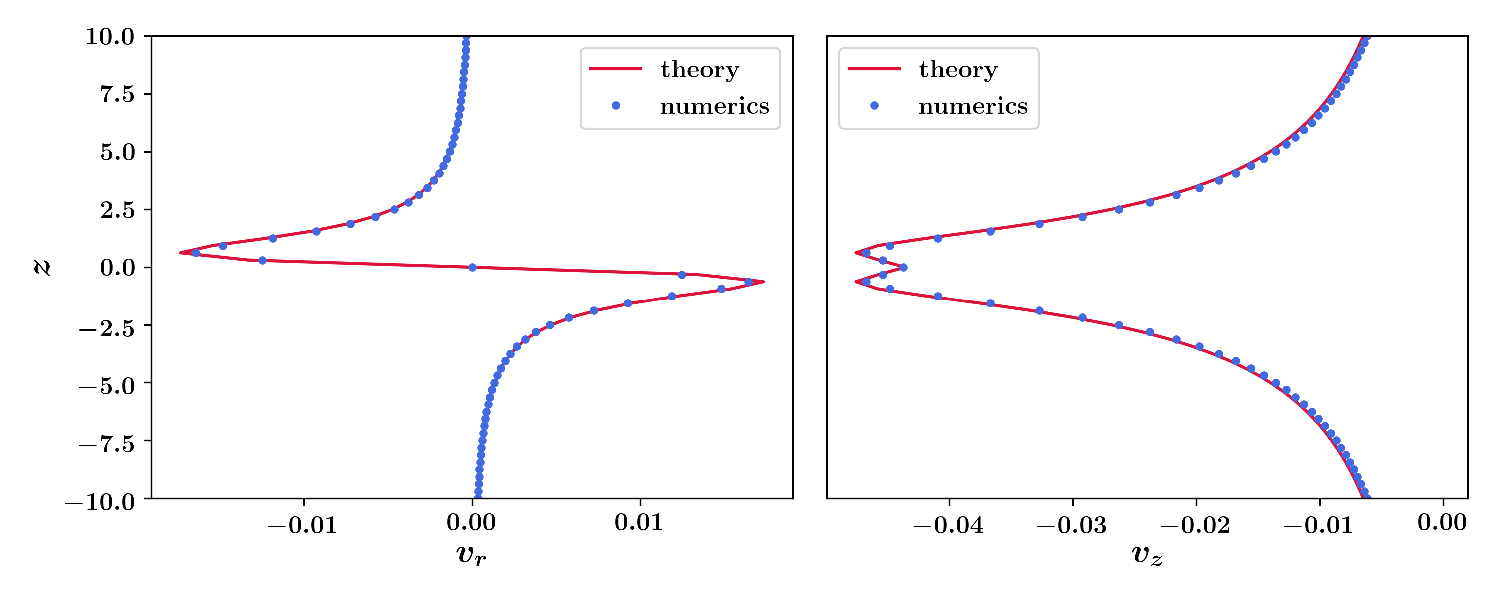}
\caption{\label{fig:num_validation}A direct comparison of the theoretical Stokeslet solution in Eq.~\ref{eq:appstokesnorm} with the numerical solution obtained using the method in Section~\ref{sec:numerics}. We plot the solution for $F_z = 1$ and $\mu = 1$, and $x = y = 0.626$. For the numerical scheme, the spacing in Fourier space is $\delta q = 0.07$ and the maximum wavenumber is $Q = 10$.
}
\label{fig:numvalid}
\end{center}
\end{figure*}

\begin{figure*}
 \begin{center}
 \leavevmode
\includegraphics[width=135mm]{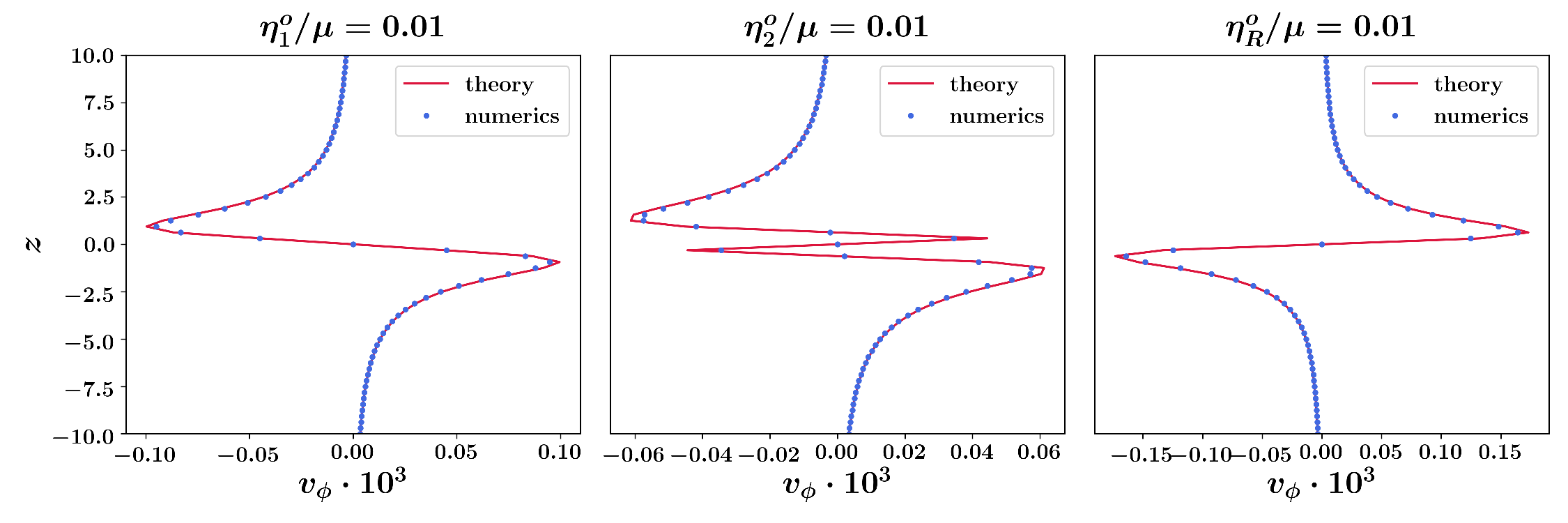} 
\caption{\label{fig:num_validation}A direct comparison of the theoretical Stokeslet solution in the presence of odd viscosity coefficients $\eta_1^o, \eta_2^o$, and $\eta_R^o$ (Eqs.~\ref{eq:eta1}, \ref{eq:eta2}, \ref{eq:etaR}, respectively) with the numerical solution obtained using the method in Section~\ref{sec:numerics}. Each panel plots the solution for one of the odd viscosities in the perturbative regime ($\eta^o_{\alpha}/\mu = 0.01$) with $F_z = 1$, $\mu = 1$,  and $x = y = 0.626$. For the numerical scheme, the spacing in Fourier space is $\delta q = 0.07$ and the maximum wavenumber is $Q = 10$.
}
\label{fig:numvalidodd}
\end{center}
\end{figure*}

The viscosity coefficients which give rise to a non-zero azimuthal component to the flow are $\eta_R^o$, $\eta^e_{Q,2}$, $\eta^o_{Q,2}$, $\eta^e_{Q,3}$, $\eta^o_{Q,3}$, $\eta_1^o$, $\eta_2^o$. We validate the numerical method in Section \ref{sec:numerics} by solving for the standard Stokeslet velocity field given in Eq. \ref{eq:appstokesnorm}. Figs.~\ref{fig:numvalid} and \ref{fig:numvalidodd} demonstrate the agreement between theory and numerics for a slice of the velocity field without and with the addition of odd viscosity, respectively.

\section{Stokeslet: exact solution for $\eta_1^o = - 2\eta_2^o$}
\label{app:exact}

Here, we provide additional details for the odd viscous Stokeslet calculation performed in Section \ref{sec:exact}.
For reference, the solution to the Stokeslet flow $
\bm{v}_{\text{Stokes},0}$ with $\bm{f} = -\bm{\hat{z}} F_z \delta^3(\bm{x})$ in a standard, isotropic fluid is given by
\begin{equation}
    \bm{v}_{\text{Stokes},0} = v_{r,0} \bm{\hat{r}} + v_{\phi,0} \bm{\hat{\phi}} + v_{r,0} \bm{\hat{z}} 
    \label{eq:appstokesnorm}
\end{equation}
where
\begin{align}
    v_{r,0}(\sphr,\theta) &= -\frac{F_z} {8 \pi \mu}\frac{\sin{\theta}\cos{\theta}}{\sphr} \nonumber\\
    v_{\phi,0}(\sphr,\theta)&= 0 \nonumber\\
    v_{z,0}(\sphr,\theta)&=-\frac{F_z} {16 \pi \mu}\frac{3 + \cos{2\theta}}{\sphr} \nonumber
\end{align}
and the pressure
\begin{equation}
    p_0(\sphr,\theta) = -\frac{F_z} {4 \pi}\frac{\cos{\theta}}{\sphr^2}
\end{equation}
Note the absence of an azimuthal component to the flow in the velocity field.

Starting from Eqs.~\ref{eq:vfourier_etas}-\ref{eq:pfourier_etas}, we consider the special case where the two shear odd viscosities satisfy the relation $\eta_1^o = -2\eta_2^o$. This simplification reduces the fields in Fourier space to
\begin{align}
\bm{\hat{v}(\bm{q})}
&=
\frac{F_z}{(q_{\perp}^2 + q_z^2)(\mu^2(q_{\perp}^2 + q_z^2)+(\eta_2^o)^2 q_z^2)}
\begin{bmatrix}
    q_z(\eta_2^o q_y + \mu q_x)\\
    \\
    q_z(-\eta_2^o q_x+ \mu q_y)\\
    \\
     -\mu q_{\perp}^2\\
\end{bmatrix}\\
\hat{p}(\bm{q})
&=
\frac{iF_z q_z[\eta_2^o(2 q_{\perp}^2 + q_z^2) + \mu^2(q_{\perp}^2 + q_z^2)]}{(q_{\perp}^2 + q_z^2)(\mu^2(q_{\perp}^2 + q_z^2)+(\eta_2^o)^2 q_z^2)}
\end{align}

To find the real space solution, we apply Eq.~\ref{eq:realspace} to the velocity and pressure fields above. Let us demonstrate the general integration method on the $\bm{\hat{x}}$-component of the velocity.

Parameterizing $\bm{q_{\perp}}$ in polar coordinates ($q_{\perp}, q_{\phi}$), we write $q_x = q_{\perp}\cos{q_{\phi}}, q_y = q_{\perp}\sin{q_{\phi}}$ and $d^2\bm{q_{\perp}} = q_{\perp}dq_{\perp}dq_{\phi}$.
Then,
\begin{align}
    \bm{{v}_{x}}(r,z) =\frac{ F_z}{(2\pi)^3}\int_0^{\infty} \dd q_{\perp}  q^2_{\perp} \int_{0}^{2\pi}&\dd q_{\phi}\,(\eta_2^o\sin{q_{\phi}} + \mu \cos{q_{\phi}})\,  e^{i q_{\perp}r \cos{(q_{\phi} - \phi})} \nonumber\\&\cdot \int_{-\infty}^{\infty}\dd q_z
    \frac{q_z e^{i q_z z}}{(q_{\perp}^2 + q_z^2)(\mu^2(q_{\perp}^2 + q_z^2)+(\eta_2^o)^2 q_z^2)}
\end{align}
The integral over $q_z$ can be taken as a contour integral in the complex plane and computed using the residue theorem. The integrand has four poles along the imaginary axis at 
\begin{align}
    q_z = \pm i |q_{\perp}|,\ q_z = \pm \frac{i\mu |q_{\perp}|}{\sqrt{\mu^2 + (\eta_2^o)^2}}
\end{align}

Then, say, for $z > 0$, we integrate over a semi-circle in the upper-half plane to find
\begin{align}
    \int_{-\infty}^{\infty} \dd q_z
    \frac{q_z e^{i q_z z}}{(q_{\perp}^2 + q_z^2)(\mu^2(q_{\perp}^2 + q_z^2)+(\eta_2^o)^2 q_z^2)}&= -\frac{i \pi\left(e^{-|q_{\perp}|z} - e^{-\frac{\mu |q_{\perp}|z}{\sqrt{(\eta_2^o)^2+\mu^2}}}\right)}{(\eta_2^o)^2 q_{\perp}^2}
\end{align}
The remaining integrals over $q_{\phi}$ and $q_{\perp}$ are straightforward, and can be computed using Mathematica or by using an integral table. Integrating over the angular part yields a Bessel function of the first kind, $J_1(q_{\perp} r)$, and the final result is given by
\begin{align}
    \bm{v_x}(r,z)=-\frac{F_z(\mu \cos{\phi} + \eta_2^o \sin{\phi})}{4\pi (\eta_2^o)^2}\frac{z}{r}\left(\frac{1}{\sqrt{r^2 + z^2}} - \frac{\mu}{\sqrt{(\eta_2^o)^2r^2 + \mu^2(r^2 + z^2)}}\right)
\end{align}

Repeating this calculation for the remaining velocity components and pressure field, rewriting in spherical coordinates and in terms of $\gamma = \eta_2^o/\mu$, we arrive at solutions given in Eqs.~\ref{eq:str}-\ref{eq:stp} in the main text.

\begin{figure}
 \begin{center}
\leavevmode
\includegraphics[width=135mm]{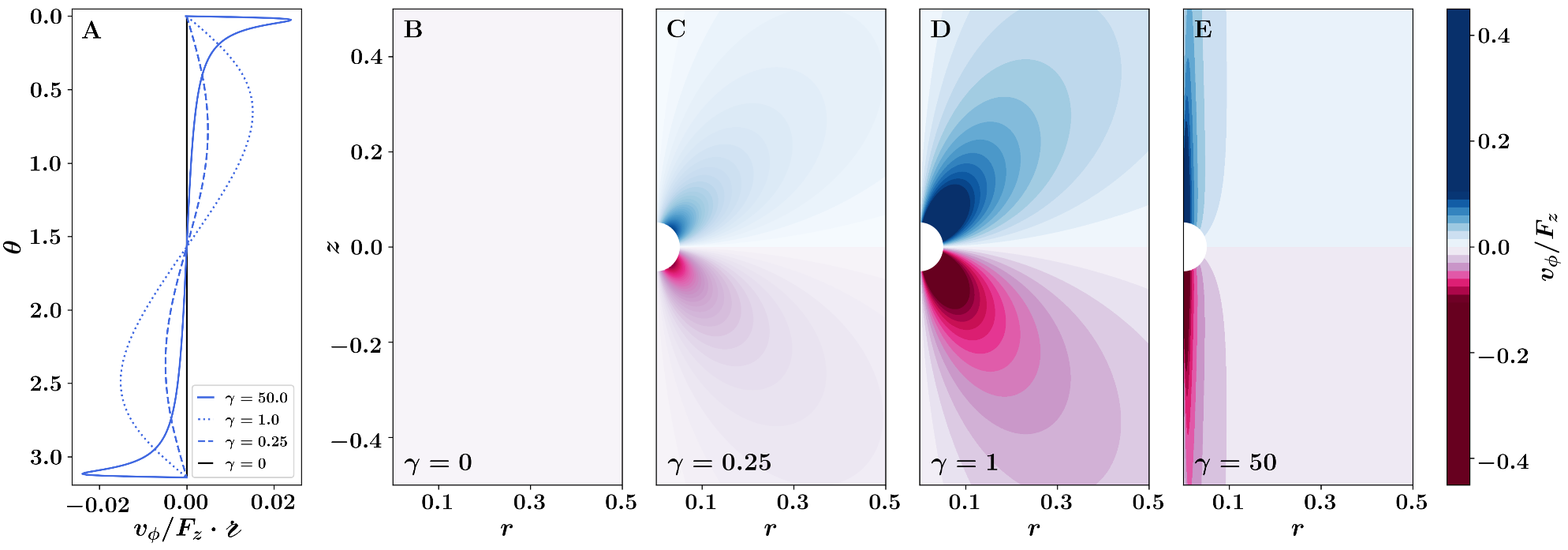} 
\caption{The structure of the azimuthal velocity component of the Stokeslet for a range of $\gamma = \eta^o/\mu$ values. A. The angular dependence of $v_{\phi}$ on $\theta$. As $\gamma$ is increased, the lobes in the azimuthal component become more pronounced, and swing out to approach the $z$-axis in the limit $\gamma \to \infty$. B-E. The azimuthal component visualized on the $r$-$z$ plane for the same $\gamma$ values as in panel A. Note the migration of the lobes as $\gamma$ is increased.}
\label{fig:appexact}
\end{center}
\end{figure}

The exact solution allows us to visualize the Stokeslet flow for a range of $\gamma$ values. In Fig.~\ref{fig:appexact}A, we plot the angular dependence of the azimuthal component of the velocity field. As $\gamma$ is increased from zero, the solution develops two lobes of opposite sign above and below the $z = 0$ plane. For high values of $\gamma$, the lobes migrate to the $z$-axis and grow in magnitude, diverging in the $\gamma \to \infty$ limit. Corresponding contour plots on the $r$-$z$ plane are found in Fig.~\ref{fig:appexact}B-E.

\begin{figure}
 \begin{center}
\leavevmode
\includegraphics[width=135mm]{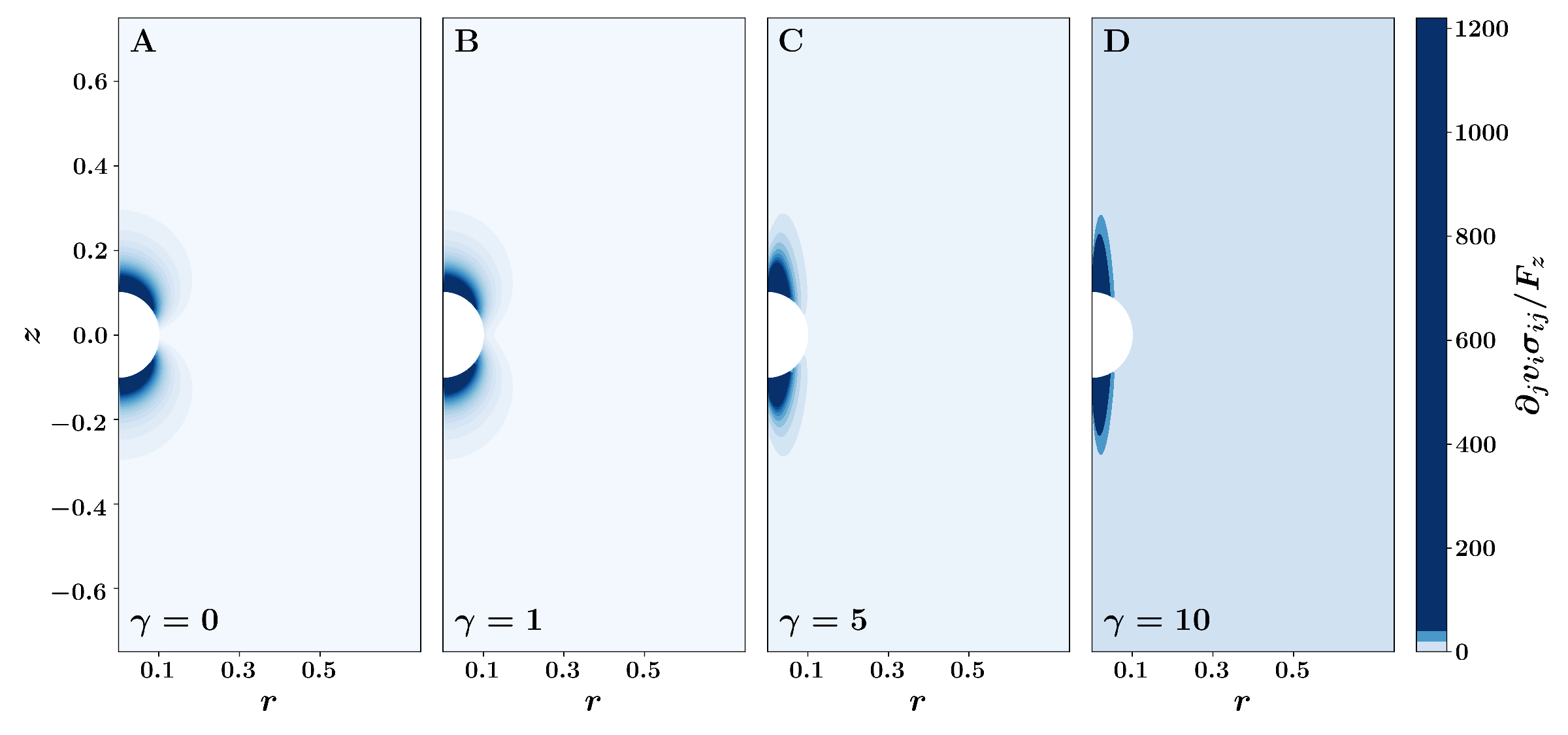} 
\caption{The energy dissipation rate of the Stokeslet flow for a range of $\gamma = \eta^o/\mu$ values, visualized on the $r$-$z$ plane. For small $\gamma$, the dissipation rate is not affected by the addition of odd viscosity. As $\gamma$ is significantly increased, the regions of high rate are concentrated near the lobes of the azimuthal flow (see Fig.~\ref{fig:appexact}).}
\label{fig:appdissipation}
\end{center}
\end{figure}

As discussed in Appendix~\ref{app:energy}, the anti-symmetric viscosity does not contribute to energy dissipation. It does, however, change the flow, so the energy dissipated by the Stokeslet in the presence of a non-zero $\gamma$ does differ from standard Stokeslet dissipation. In Fig.~\ref{fig:appdissipation}, we show contour plots of $\dot{w} = \sigma_{ij} (\partial_j v_i)$ for a range of $\gamma$ values. Although the contribution to the dissipation vanishes at first-order in $\eta^o_2/\mu$, for larger values of $\gamma$, the regions of high dissipation rate are concentrated near the lobes of the azimuthal component of the flow.

We note that the limit $\eta^o \equiv \eta_{1}^o - 2 \eta_2^o$ and $\mu \equiv \mu_1 =\mu_2 = \mu_3 $ has a simple geometric interpretation. Restricting ourselves to to the shear subspace, the matrix $\eta^{AB}$ has the form $\eta^{AB} = \eta^o L_z^{AB} + \mu \delta^{AB} $, where $L_z^{AB}$ is the generator of rotations about the $\bm{\hat{z}}$ axis. In this sense, the $\mu$ seeks to cause contractions in shear space, and $L_z$ generates azimuthal rotations in shear space. 

\section{Stokeslet: perturbative solution for small $\eta_1^o, \eta_2^o$, and $\eta_R^o$}
\label{app:stokeslet}

In this section, we present additional details for the calculations performed in Section \ref{sec:stokeslet}. Here, we assume that the odd viscosity is small as compared to the even viscosity, and consider each coefficient separately.

Let us demonstrate the method with $\eta_1^o$.
Setting $\eta_2^o = 0$, and expressing the viscosity ratio as $\epsilon_1 = \eta^o_1/\mu \ll 1$, the velocity field in Fourier space (Eq.~\ref{eq:vfourier_etas}) reduces to
\begin{align}
\bm{\hat{v}(\bm{q})}
&=
-\frac{F_z}{\mu}\frac{1}{(q_{\perp}^2 + q_z^2)^3+\epsilon_1^2q_{\perp}^4 q_z^2}
\begin{bmatrix}
    q_z(\epsilon_1 q_y q_{\perp}^2 -q_x (q_{\perp}^2 + q_z^2))\\
    \\
    q_z(-\epsilon_1 q_x q_{\perp}^2 -q_y (q_{\perp}^2 + q_z^2))\\
    \\
     q_{\perp}^2(q_{\perp}^2 + q_z^2)\\
\end{bmatrix}
\end{align}

The zeroth order field is simply the standard Stokeslet solution,
\begin{align}
\bm{\hat{v}_0(\bm{q})}
&=
-\frac{F_z}{\mu}\frac{1}{(q_{\perp}^2 + q_z^2)^3}
\begin{bmatrix}
    -q_z q_x (q_{\perp}^2 + q_z^2)\\
    \\
    -q_z q_y (q_{\perp}^2 + q_z^2)\\
    \\
     q_{\perp}^2(q_{\perp}^2 + q_z^2)\\
\end{bmatrix}
\end{align}
Meanwhile, the leading order correction is linear in $\epsilon_1$, and is given by
\begin{align}
\bm{\hat{v}_1(\bm{q})}
&=
-\epsilon_1 \frac{F_z}{\mu}\frac{1}{(q_{\perp}^2 + q_z^2)^3}
\begin{bmatrix}
    q_z q_y q_{\perp}^2 \\
    \\
    -q_z q_x q_{\perp}^2 \\
    \\
    0\\
\end{bmatrix}
\end{align}

To obtain the real space solution, we once again apply Eq.~\ref{eq:realspace}, and integrate as delineated in Appendix \ref{app:exact}. In this case, the poles for the integral over $q_z$ are given by $q_z = \pm i \abs{q_{\perp}}$. With this method, we find the emergence of an azimuthal component to the flow (Eq.~\ref{eq:eta1}), and no correction to the pressure at first order in $\epsilon_1$.

We proceed similarly for the remaining coefficients. For $\epsilon_2 = \eta_2^o/\mu$, we set $\eta_1^o = 0$ and again expand the denominator in Eq.~\ref{eq:vfourier_etas}. For the rotational viscosities, we instead expand the expressions in Eqs.~\ref{eq:vfourier_rot}-\ref{eq:fourier_rot_denom}.

Unlike $\eta_{R,1}$ and $\eta_R^o$, the rotational viscosity $\eta_{R,2}$ does not affect the Stokeslet flow. To see this, let us begin by inspecting the Stokes equation at first order in $\epsilon_{R,2} = \eta_{R,2}/\mu$. Writing $\bm{v} = \bm{v_0} + \epsilon_{R,2}\bm{v_1}, P = P_0 + \epsilon_{R,2}P_1$, we find the first order equation to be 
\begin{align}
    0 = -\nabla P_1/\mu + \Delta \bm{v_1} + \Delta_{\eta_{R,2}} \bm{v_0}
\end{align}
Here $\Delta_{\eta_{R,2}}$ in cylindrical coordinates is 
\begin{align}
    \Delta_{\eta_{R,2}}\bm{v}
    =
    \begin{bmatrix}
        0\\
        \\
        -\partial_r \omega_z\\
        \\
        0
    \end{bmatrix}
    \label{eq:etaR2}
\end{align}
where $\omega_z$ is the $\bm{\hat{z}}$-component of the vorticity. From Eq.~\ref{eq:appstokesnorm}, we find $\omega_{z, 0} = 0$, so the last term in Eq.~\ref{eq:etaR2} vanishes. Thus, the trivial solution $\bm{v_1} = 0, P_1 = 0$ satisfies Eq.~\ref{eq:etaR2}. In fact, the flow is unaffected at all orders; at order $m$ in $\epsilon_{R,2}$, the term $\Delta_{\eta_{R,2}}\bm{v_{m-1}}$ is zero since $\bm{v_{m-1}} = 0$, so the trivial solution always satisfies the equation.

As an example, we compute the full Oseen tensor in the limit $\eta^o \equiv \eta_2^o = -\eta_1^o/2$, in the perturbative regime $\epsilon \equiv \eta^o/\mu \ll 1$. We find
\begin{align}
    \label{perturbative_oseen_tensor}
    G = 
    \frac{1}{8 \pi \mu \sphr^3}
    \begin{bmatrix}
    2x^2 + y^2 + z^2 && xy && xz\\
    xy && x^2 + 2y^2 + z^2 && yz \\
    xz && yz && x^2 + y^2 + 2z^2
    \end{bmatrix}
    +\\ \notag
    \frac{\epsilon}{8 \pi \mu \sphr^3}
    \begin{bmatrix}
    0 && x^2 + y^2 && yz\\
    -(x^2 + y^2) && 0 && -xz\\
    -yz && xz && 0
    \end{bmatrix}.
\end{align}

\section{Viscous flow past a sphere}
\label{app:sphere}
Here, we provide additional details for the calculation of odd viscous flow past a sphere performed in Section \ref{sec:sphere}.
For reference, in a standard isotropic fluid, the velocity and pressure fields for viscous flow past a sphere in the $\bm{\hat{z}}$-direction are given by
\begin{align}
    v_{r,0}(\sphr, \theta) &= -\frac{3aU\sin{2\theta}}{8\sphr} + \frac{3a^3 U \sin{2\theta}}{8\sphr^3} \\
    v_{\phi,0}(\sphr,\theta) &= 0\\
    v_{z,0}(\sphr,\theta) &= U - \frac{3aU(3 + \cos{2\theta})}{8\sphr} + \frac{a^3U(1 + 3\cos{2\theta})}{8\sphr^3}\\
    p_0(\sphr,\theta) &= -\frac{3aU\mu \cos{\theta}}{2\sphr^2} + \mbox{const}
    \label{eq:appnormsphere}
\end{align}
Note the absence of an azimuthal component to the flow.

As described in Section \ref{sec:sphere}, we work in a perturbative regime and assume that the pressure correction vanishes at linear order. In this case, the Stokes flow equation reduces to the Poisson equation for the first order velocity field,
\begin{align}
    \Delta \bm{v_1} = - \Delta_{\alpha} \bm{v_0} 
    \label{eq:appvecpoisson}
\end{align}
In cartesian coordinates, the Green function is given by
\begin{align*}
\mathbbm{G}(\bm{x},\bm{x'}) = -\frac{1}{|\bm{x}-\bm{x'}|} \equiv -\frac{1}{R}
\end{align*}
For ease of dealing with the boundary condition on the sphere, 
we work in spherical coordinates and we obtain the vector Laplacian in spherical coordinates as follows.  Writing $F = \Delta_{\alpha} \bm{v_0}$, the solution to Eq. \ref{eq:appvecpoisson} in Cartesian coordinates can be written as 
\begin{align}
    \bm{v_1}(\bm{x}) = -\int \dd^3 x' \mathbbm{G}(\bm{x, x'}) \bm{F}(\bm{x'})
\end{align}
To convert a cartesian vector to spherical coordinates, we apply the matrix $\mathsfbi{T}$,
\begin{align}
\begin{bmatrix}
    \hat{\bm{\sphr}}  \\
    \hat{\bm{\phi}} \\
    \hat{\bm{\theta}} \\
\end{bmatrix}
=
\begin{bmatrix}
    \cos{\phi}\sin{\theta} && \sin{\phi}\sin{\theta} && \cos{\theta}\\
    -\sin{\phi} && \cos{\phi} && 0\\
    \cos{\phi}\cos{\theta} && \sin{\phi}\cos{\theta} && -\sin{\theta}\\
\end{bmatrix}
\begin{bmatrix}
    \hat{\bm{x}}  \\
    \hat{\bm{y}} \\
    \hat{\bm{z}} \\
\end{bmatrix}
\label{eq:appsphconv}
\end{align}
Denoting vectors in spherical coordinates with a tilde, we write
\begin{align*}
    \widetilde{\bm{v_1}}(\bm{x}) &= -\int \dd^3 x' \mathsfbi{T}(\bm{x}) \mathbbm{G}(\bm{x, x'}) \bm{F}(\bm{x'})\\
    &=-\int \dd^3 x' \mathsfbi{T}(\bm{x}) \mathbbm{G}(\bm{x, x'}) \mathsfbi{T}^{-1}(\bm{x'})\widetilde{\bm{F}}(\bm{x'})\\
    &=-\int \dd^3 x' \widetilde{\mathbbm{G}}(\bm{x, x'}) \widetilde{\bm{F}}(\bm{x'})
\end{align*}
where $\widetilde{\mathbbm{G}}(\bm{x, x'}) \equiv \mathsfbi{T}(\bm{x}) \mathbbm{G}(\bm{x, x'}) \mathsfbi{T}^{-1}$.
We find that in spherical coordinates, $\widetilde{\mathbbm{G}}(\bm{x, x'})
$ is given by
\begin{small}
\begin{align}
-\frac{1}{R}
\begin{bmatrix}
   \cos{\theta}\cos{\theta'} + \cos{(\phi-\phi')}\sin{\theta}\sin{\theta'} && \sin{\theta}\sin{(\phi-\phi')} && \sin{\theta}\cos{\theta'}\cos{(\phi - \phi')} - \cos{\theta}\sin{\theta'}\\
   -\sin{\theta'}\sin{(\phi - \phi')} && \cos{(\phi - \phi')} && -\cos{\theta'}\sin{(\phi - \phi')}\\
   -\cos{\theta'} \sin{\theta} + \cos{\theta}\sin{\theta'}\cos{(\phi - \phi')}&& \cos{\theta} \sin{(\phi - \phi')} && \cos{\theta}\cos{\theta'}\cos{(\phi - \phi')}+\sin{\theta}\sin{\theta'}
\end{bmatrix}
\label{eq:sphG}
\end{align}
\end{small}
Since this Green function is not diagonal, the different source components mix. For the odd viscosities, we have $\bm{F} = F_{\phi}\bm{\hat{\phi}}$, so 
\begin{align}
    \widetilde{\mathbbm{G}}(\bm{x, x'})\widetilde{\bm{F}}(\bm{x'}) = 
    -\frac{1}{R}
    \begin{bmatrix}
        F_{\phi} \sin{\theta}\sin{(\phi - \phi')}\\
        F_{\phi} \cos{(\phi-\phi')}\\
        F_{\phi}\cos{\theta}\sin{(\phi - \phi')}
    \end{bmatrix}
    \label{eq:appGF}
\end{align}

The odd source terms, written in spherical coordinates, are given below, for $\eta_1^o, \eta_2^o$, and $\eta_R$, respectively: 
\begin{align}
    F_{\phi}(\sphr, \theta) &= \frac{3aU(5a^2 - 9\sphr^2)}{16\sphr^5}\sin{2\theta} + \frac{15aU(7a^2 - 3\sphr^2)}{16\sphr^5}\cos{2\theta}\sin{2\theta}\\
    F_{\phi}(\sphr, \theta) &= \frac{3aU(5a^2 - 3\sphr^2)}{8\sphr^5}\sin{2\theta} + \frac{15aU(7a^2 - 3\sphr^2)}{8\sphr^5}\cos{2\theta}\sin{2\theta}\\
    F_{\phi}(\sphr, \theta) &= \frac{9aU}{4\sphr^3}\sin{2\theta}
    \label{eq:appspheresource}
\end{align}

We absorb the spherical corrections to the Green function into the source, and expand the Cartesian Green function in spherical harmonics,
\begin{align}
    \mathbbm{G}(\bm{x, x'}) =- \sum_{\ell = 0}^{\infty} \sum_{m = -\ell}^{\ell} \frac{1}{2\ell+1} \frac{\sphr^\ell_{<}}{\sphr^{\ell+1}_{>}} Y_\ell^m (\theta, \phi) \bar{Y}_\ell^m (\theta', \phi')
\end{align}
This Green function, however, does not satisfy the boundary condition on the sphere (no-slip); it only guarantees a well-behaving solution at infinity. Instead, we need to use the Dirichlet Green function, where we can impose $\bm{v_1}(\sphr=a, \theta) = 0$. The relevant Dirichlet Green function is known: a direct electrostatics analogy for this problem is a conducting spherical cavity of radius $a$ with a point charge placed at $\bm{x'}$ and a vanishing potential on the surface of the sphere. The Green function for this problem can be found using the method of images \citep{jackson1999classical}, and is given by
\begin{align}
    \mathbbm{G}_D(\bm{x, x'}) &=- \frac{1}{\abs{\bm{x} - \bm{x'}}} + \frac{a}{\sphr'\abs{\bm{x} - \frac{a^2}{\sphr'^2}\bm{x'}}}\\
    &=-\sum_{\ell = 0}^{\infty} \sum_{m = -\ell}^{\ell} \frac{1}{2\ell+1} \left[\frac{\sphr^\ell_{<}}{\sphr^{\ell+1}_{>}} - \frac{1}{a}\left(\frac{a^2}{\sphr\sphr'}\right)^{\ell+1}\right] Y_\ell^m (\theta, \phi) \bar{Y}_\ell^m (\theta', \phi')
    \label{eq:GD}
\end{align}
where
\begin{align}
    \frac{\sphr^\ell_{<}}{\sphr^{\ell+1}_{>}} - \frac{1}{a}\left(\frac{a^2}{\sphr\sphr'}\right)^{\ell+1} = \begin{cases}
    \frac{1}{\sphr'^{\ell+1}}\left(\sphr^\ell - \frac{a^{2\ell+1}}{\sphr^{\ell+1}} \right), & \sphr <\sphr'\\
    \frac{1}{\sphr^{\ell+1}}\left(\sphr'^\ell - \frac{a^{2\ell+1}}{\sphr'^{\ell+1}} \right), & \sphr >\sphr'\\
    \end{cases}
\end{align}
Then, we evaluate the integral below with the Dirichlet Green function
\begin{align}
    \widetilde{\bm{v_1}}(\bm{x}) &= -\int \dd^3 x' \widetilde{\mathbbm{G}_D}(\bm{x, x'}) \widetilde{\bm{F}}(\bm{x'})
    \label{eq:appsphereint}
\end{align}
by using the spherical harmonics expansion in Eq. \ref{eq:GD}, and find the velocity fields given in Eqs.~\ref{eq:sphere}.

\section{Viscous flow past a bubble}
\label{app:bubble}

Here, we provide additional details for the calculation of odd viscous flow past a spherical bubble performed in Section \ref{sec:bubble}. In a standard, isotropic fluid, the velocity field for the flow outside the bubble is 

\begin{align}
    v_r(\sphr, \theta) &= -\frac{5aU \cos{\theta}\sin{\theta}}{8\sphr} + \frac{3a^3U\cos{\theta}\sin{\theta}}{8\sphr^3}\\
    v_{\phi} (\sphr, \theta) &= 0\\
    v_z(\sphr, \theta) &= U - \frac{5aU(3 + \cos{2\theta})}{16 \sphr} + \frac{a^3U(1 + 3\cos{2\theta})}{16 \sphr^3}
\end{align}

Inside the bubble, the fluid forms Hill's spherical vortex, given by
\begin{align}
    v_r(\sphr, \theta) &= -\frac{U}{4a^2}\sphr^2 \cos{\theta}\sin{\theta}\\
    v_{\phi}(\sphr, \theta) &= 0\\
    v_z(\sphr, \theta) &= -\frac{U}{8a^2}(2a^2 - 3\sphr^2 + \sphr^2 \cos{2\theta})
\end{align}

As in the previous problems we consider, the standard flow is two-dimensional, with no azimuthal component. To evaluate the effect of odd viscosity on the bubble flow, we work in a perturbative regime, with $\eta^o \ll \mu$. 

Outside the bubble, the problem is remarkably similar to flow past a sphere, without the requirement of the no-slip boundary conditions on the surface. We again look for a solution of the vector Poisson equation
\begin{align}
    \Delta \bm{v_1} = - \Delta_{\alpha} \bm{v_0} 
    \label{eq:appvecpoisson2}
\end{align}

As in the case of the sphere, the odd source terms, $\bm{F} = \Delta_{\alpha} \bm{v_0}$, only have an azimuthal component, taking the form

\begin{align}
    F_{\phi}(\sphr, \theta) &= -\frac{15aU (a^2 - 3\sphr^2)}{32 \sphr^5}\sin{2\theta} - \frac{15aU(7a^2 - 5\sphr^2)}{32\sphr^5}\cos{2\theta}\sin{2\theta}\\
    F_{\phi}(\sphr, \theta) &= -\frac{15aU(a^2 - \sphr^2)}{16\sphr^5}\sin{2\theta} - \frac{15aU(7a^2 - 5\sphr^2)}{16\sphr^5}\cos{2\theta}\sin{2\theta}\\
    F_{\phi}(\sphr, \theta) &= -\frac{15aU}{8\sphr^3}\sin{2\theta}
\end{align}

Following the calculation leading to Eq.~\ref{eq:appGF}, we absorb the spherical corrections to the Green function into the source, and expand the Cartesian Green function in spherical harmonics. This time, the standard expansion suffices \citep{jackson1999classical},
\begin{align}
    \mathbbm{G}(\bm{x, x'}) &=- \frac{1}{\abs{\bm{x} - \bm{x'}}} \\
    &=-\sum_{\ell = 0}^{\infty} \sum_{m = -\ell}^{\ell} \frac{1}{2\ell+1} \left[\frac{\sphr^\ell_{<}}{\sphr^{\ell+1}_{>}} \right] Y_\ell^m (\theta, \phi) \bar{Y}_\ell^m (\theta', \phi')
    \label{eq:G}
\end{align}
where
\begin{align}
    \frac{\sphr^\ell_{<}}{\sphr^{\ell+1}_{>}}
    = \begin{cases}
    \frac{\sphr^\ell}{\sphr'^{\ell+1}}, & \sphr <\sphr'\\
    \frac{\sphr'^\ell}{\sphr^{\ell+1}}, & \sphr >\sphr'\\
    \end{cases}
\end{align}
Evaluating the integral in Eq.~\ref{eq:appsphereint} with this Green function, we find the velocity fields given in Eqs.~\ref{eq:bubbleout}.

Let us now consider the flow inside the bubble. Once again assuming $p_1 = 0$, we are left with Eq.~\ref{eq:appvecpoisson2}. In this case, however, the term $\Delta_{\alpha} \bm{v_0} = 0$, so Eq.~\ref{eq:appvecpoisson2} reduces to the vector Laplace equation,
\begin{equation}
    \Delta \bm{v_1} = 0
    \label{appveclaplace}
\end{equation}
with the boundary condition $\bm{v_{1, \text{out}}}(a, \theta) = \bm{v_{1, \text{in}}} (a, \theta)$. 
The velocity on the boundary in spherical coordinates is given by
\begin{align}
    v_{\phi}(a, \theta) &= -\frac{1}{7} U \cos{\theta}\sin{\theta}\\
    v_{\phi}(a, \theta) &= -\frac{1}{28} U \cos{\theta}\sin{\theta} \\
    v_{\phi}(a, \theta) &= \frac{1}{4} U \cos{\theta}\sin{\theta}
\end{align}
for $\eta_1^o, \eta_2^o$, and $\eta_R^o$, respectively.

Here, as in the case of the sphere, we can make an analogy with electrostatics. Our setup is the vector version of the following situation: a spherical cavity, with no charge inside, but a potential specified to be some function on the surface of the cavity. This is known as the ``Dirichlet problem", and can be solved with the use of the Dirichlet Green function $\mathbbm{G}_D$ from Eq.~\ref{eq:GD}.
In cartesian coordinates, the solution to our Dirichlet problem is given by
\begin{equation}
    \bm{v_1(\bm{x})} = -\int \dd^2 x' \bm{\hat{\sphr}'}\cdot \nabla_{\bm{x}'}\mathbbm{G}_D(\bm{x'}, \bm{x}) \bm{v_{1, \text{out}}}(a, \theta')
\end{equation}
Again, it is convenient to work in spherical coordinates. Denoting vectors in spherical coordinates with a tilde, we transform
\begin{align*}
    \widetilde{\bm{v_1}}(\bm{x}) &= - \int \dd^2 x' \mathsfbi{T}(\bm{x}) \bm{\hat{\sphr}'}\cdot \nabla_{\bm{x}'}\mathbbm{G}_D(\bm{x'}, \bm{x}) \bm{v_{1, \text{out}}}(a, \theta')\\
    &=- \int \dd^2 x' \mathsfbi{T}(\bm{x}) \bm{\hat{\sphr}'}\cdot \nabla_{\bm{x}'}\mathbbm{G}_D(\bm{x'}, \bm{x}) \mathsfbi{T}^{-1}(\bm{x'})\widetilde{\bm{v}}_{1, \text{out}}(a, \theta')\\
    &=- \int \dd^2 x' \mathsfbi{T}(\bm{x}) \partial_{\sphr'}\mathbbm{G}_D(\bm{x'}, \bm{x}) \mathsfbi{T}^{-1}(\bm{x'})\widetilde{\bm{v}}_{1, \text{out}}(a, \theta')\\
    &=- \int \dd^2 x' \partial_{\sphr'} [\mathsfbi{T}(\bm{x}) \mathbbm{G}_D(\bm{x'}, \bm{x}) \mathsfbi{T}^{-1}(\bm{x'})]\widetilde{\bm{v}}_{1, \text{out}}(a, \theta')\\
    &= - \int \dd^2 x' \partial_{\sphr'}\widetilde{\mathbbm{G}_D}(\bm{x'}, \bm{x}) \widetilde{\bm{v}}_{1, \text{out}}(a, \theta')
\end{align*}
where $\widetilde{\mathbbm{G}_D}(\bm{x, x'}) \equiv \mathsfbi{T}(\bm{x}) \mathbbm{G}_D(\bm{x, x'}) \mathsfbi{T}^{-1}(\bm{x'})$, and $\mathsfbi{T}$ is defined in Eq.~\ref{eq:appsphconv}.

Taking into account the mixing of the source components in Eq.~\ref{eq:sphG}, we compute this integral by expanding the Dirichlet Green function in spherical harmonics, as in Eq.~\ref{eq:GD}. The relevant expansion for this ``interior" problem \citep{jackson1999classical} is given by
\begin{align}
    \mathbbm{G}_D(\bm{x, x'}) 
    &=-\sum_{\ell = 0}^{\infty} \sum_{m = -\ell}^{\ell} \frac{1}{2\ell+1} \left[\frac{\sphr^\ell_{<}}{\sphr^{\ell+1}_{>}} - \frac{1}{a}\left(\frac{\sphr \sphr'}{a^2}\right)^\ell\right] Y_\ell^m (\theta, \phi) \bar{Y}_\ell^m (\theta', \phi')
    \label{eq:GD}
\end{align}
where
\begin{align}
    \frac{\sphr^\ell_{<}}{\sphr^{\ell+1}_{>}} - \frac{1}{a}\left(\frac{\sphr \sphr'}{a^2}\right)^{\ell}
    = \begin{cases}
    \frac{\sphr^\ell}{\sphr'^{\ell+1}} -  \frac{1}{a}\left(\frac{\sphr \sphr'}{a^2}\right)^\ell, & \sphr <\sphr'\\
    \frac{\sphr'^\ell}{\sphr^{\ell+1}} -  \frac{1}{a}\left(\frac{\sphr \sphr'}{a^2}\right)^\ell, & \sphr >\sphr'
    \end{cases}
\end{align}

The resulting velocity fields are given in Eq.~\ref{eq:bubblein}.

\section{Supplementary Movies}
\label{app:movies}

\textbf{Supplementary Movie 1}.\\
Stokeslet solutions for a range of odd to even viscosity ratios, in the limit $\eta_1^o = -2\eta_2^o$. An external force is applied at the origin along the $-\bm{\hat{z}}$ direction. The curves are streamlines propagating from a set of initial conditions arranged in a circle. As $\eta_2^o$ is added ($\gamma = \eta_2^o/\mu > 0$), the velocity field develops an azimuthal component that changes sign across the $z = 0$ plane, where the source is located. In the limit of only $\eta_2^o$ ($\gamma \to \infty$), the familiar radial component of the flow vanishes.

\textbf{Supplementary Movie 2}.\\
Sedimentation of a cloud in an odd viscous fluid, in the limit $\eta_1^o = - 2\eta_2^o$. Each cloud is initialized with $N = 2000$ particles, where each particle is treated as a Stokeslet falling under the effect of gravity. Only particles within a few radii of the cloud are tracked. Top-down (A) and side (B) views are shown for three different values of the odd to even viscosity ratio, $\gamma = \eta_2^o/\mu$.
\textbf{Part 1.} A cloud falling in the absence of odd viscosity, $\gamma = 0$. As the cloud leaks particles from its outermost layer into a vertical tail, the center of the cloud is depleted. As a result, the cloud deforms into a torus, and subsequently breaks apart into smaller clouds.
\textbf{Part 2.} When the odd viscosity is increased to $\gamma = 1$, the particle streamlines acquire a rotating azimuthal component. For this value of $\gamma$, the cloud still forms a torus, although the later break-up event only happens in a fraction of the simulation runs.
\textbf{Part 3.} In a highly odd viscous fluid ($\gamma = 5$), the initially spherical cloud immediately deforms into an ellipsoid. The azimuthal flow within the cloud dominates the radial velocity component. In this regime, the cloud no longer forms a torus, and the instability is suppressed.

\section{Acknowledgements}
We thank Ming Han and Tom Witten. V.V.~acknowledges support from the Simons Foundation, the Complex Dynamics and Systems Program of the Army Research Office under grant W911NF-19-1-0268, and the University of Chicago Materials Research Science and Engineering Center, which is funded by the National Science Foundation under Award No.~DMR-2011854.
T.K.~and C.S.~were supported by the National Science Foundation Graduate Research Fellowship under Grant No.~1746045.
M.F. acknowledges support from a MRSEC-funded Kadanoff–Rice fellowship (DMR-2011854) and the Simons Foundation.
Some of us benefited from participation in the KITP program on Symmetry, Thermodynamics and Topology in Active Matter supported by Grant No. NSF PHY-1748958.

\bibliographystyle{jfm}

\end{document}